\documentclass[journal=tches,preprint]{iacrtrans}

\usepackage{lipsum} 
\usepackage{amssymb}
\usepackage{amsmath}
\usepackage{subcaption}
\usepackage{algorithm,algpseudocode}
\usepackage{pifont}
\usepackage{multirow}
\usepackage[flushleft]{threeparttable}

\newcommand{\cmark}{\ding{51}}
\newcommand{\xmark}{\ding{55}}

\newcommand{\corr}[1]{\textcolor{black}{#1}}

\author{Benoît Coqueret\inst{2} \and 
Mathieu Carbone\inst{2} \and
Olivier Sentieys\inst{3} \and 
Gabriel Zaid\inst{1}\inst{\thanks{Work done when the author was at Thales ITSEF.}}
}

\institute{
  CryptoExperts, Paris, France, \email{{firstname.lastname}@cryptoexperts.com} \and
  Thales ITSEF, Toulouse, France, \email{{firstname.lastname}@thalesgroup.com}
  \and
  University of Rennes, INRIA, IRISA, Rennes, France, \email{{firstname.lastname}@inria.fr}
}

\title[A Divide-and-Conquer Strategy for Hard-Label Extraction of DNN via SCA]{A Divide-and-Conquer Strategy for Hard-Label Extraction of Deep Neural Networks via Side-Channel Attacks}

\begin{document}

\maketitle


\begin{abstract}
During the past decade, Deep Neural Networks (DNNs) have proven their value across a wide variety of applications; however, despite their importance, protecting their intellectual property remains an open issue.
Recent work has successfully extracted DNNs using cryptanalytic methods in hard-label settings, showing that it is possible to copy a DNN with high fidelity, \textit{i.e.}, a high degree of similarity in \corr{correct/incorrect} output predictions which corresponds to the proportion of samples for which the extracted model produces the same prediction as the original model.
However, these methods have only been demonstrated on Multi-Layer Perceptrons (MLPs) and are sensitive to non–fully connected layers and special-case neurons.
To overcome these limitations, we base our contribution on a divide-and-conquer paradigm.
We introduce a new black-box \corr{side-channel attack} that splits the targeted DNN into several linear components, for which cryptanalytic extraction can be performed.
Building on this decomposition, we propose an end-to-end framework specifically designed for hard-label settings, not limited to fully connected layers, and robust to special-case neurons, while improving extraction fidelity.
We validate our contribution by successfully extracting all architectures previously targeted in the literature, as well as several new architectures implemented on a microcontroller unit.
These include an MLP with $1.7$ million parameters, nearly doubling the previous largest number of extracted weights, and a shortened MobileNetv1, which for the first time includes pooling layers and depthwise separable convolutions.
Our framework successfully extracts all of these DNNs with high fidelity ($88.4\%$ for MobileNetv1 and $93.2\%$ for the MLP).
Finally, we use the \corr{copied} model to generate adversarial examples and achieve near white-box performance on the victim model ($95.8\%$ and $96.7\%$ transfer rates).
\end{abstract}





\section{Introduction}
\label{introduction}
During the last decade, the number of tasks for which Deep Neural Networks (DNNs) have proven their effectiveness has steadily increased, leading to the widespread adoption of these algorithms in a large variety of fields.
From computer vision to text translation, DNNs are now everywhere, and the best models have become valuable intellectual property (IP).
At the same time, parallel efforts from hardware designers have made the deployment of DNNs on edge devices possible.
However, due to their high value, the IP of deployed models must be protected against new attacks arising from the embedded-system context.
Since the publication of the first side-channel attack against the IP of an embedded DNN \cite{Batina2019CSINR}, the number of physical attacks against DNNs has greatly increased.
Several methodologies using side-channel attacks with the objective of extracting DNN hyperparameters have been proposed \cite{Joud2023LikeAO, Gao2023DeepTheftSD}.
DNN parameters have also been targeted by physical attacks, either via side-channel analysis \cite{9300274, Joud2022API} or through fault injection \cite{Rakin2021DeepStealAM, Hector2023FaultIA}.
These types of attacks, which target the parameters of the model, aim to copy the targeted model and thus perform a model extraction attack.

Model extraction is a threat to any deployed DNN, and as such there exists a large variety of methods—and even objectives—for these attacks.
We can characterize the two main types of objectives, or adversarial goals, for model extraction using the terminology introduced in \cite{Jagielski2019HighAA}: \textit{accuracy-based model extraction} and \textit{fidelity-based model extraction}.
The former aims to obtain a substitute model with good performance on the task of the targeted model without performing the full training process.
The latter aims to \corr{retrieve the exact internal parameters (\textit{e.g.}, trainable weights) of} the targeted model in order to obtain a \corr{copy} as close as possible to the original.
The copied model can then be used to gain information about the victim’s DNN and potentially mount more powerful attacks against it.
In this study, we consider only \textit{fidelity-based model extraction}.

\paragraph{Cryptanalytic-based extraction.} 
Jagielski \textit{et al.} \cite{Jagielski2019HighAA} were the first to propose a functional methodology for fidelity-based model extraction of single-layer neural networks (NNs) using the ReLU (Rectified Linear Unit) activation function.
They exploit the gradient of the DNN to gain access to what they define as critical points.
These points correspond to inputs for which the activation value of a specific neuron is zero.
Such points are observable in DNNs using ReLU activations, as they correspond to locations where discontinuities occur in the gradient of the network.
Using this observation, they are able to extract a single-layer fully connected NN using a least-squares (LSTSQ) algorithm.
Both \cite{Carlini2020CryptanalyticEO} and \cite{Rolnick2019ReverseengineeringDR} successfully extend this result to deeper architectures, with the method proposed in \cite{Carlini2020CryptanalyticEO} achieving the highest fidelity to date.
One key limitation of the method proposed in \cite{Carlini2020CryptanalyticEO} is the extraction of neuron signs, which relies on an exhaustive search.
This information is critical, as assigning an incorrect sign \corr{leads to an unsuccessful attack (\textit{i.e.}, activated neurons become deactivated, and vice versa)}.
Canales-Martínez \textit{et al.} argue that this approach does not scale well to larger architectures \cite{Shamir2023PolynomialTC}.
They instead use crafted perturbations that activate specific neurons to infer this information \cite{Shamir2023PolynomialTC}.
While effective, all these methods rely on access to the full confidence-score output vector of the DNN, which enables gradient estimation.
This represents an idealized attacker setting, and recent works have aimed to remove this assumption in order to move toward more realistic scenarios.
Chen \textit{et al.} \cite{Chen2024HardLabelCE} recently demonstrated that it is possible to extract DNNs in hard-label settings for small networks.
This result was further extended by Carlini \textit{et al.} \cite{Carlini2024PolynomialTC}, who successfully extracted a DNN with four hidden layers using only hard-label outputs.
However, although these works no longer require confidence scores, their designs remain sensitive to non–fully connected layers and special-case neurons, as in previous methods \cite{Carlini2020CryptanalyticEO, Shamir2023PolynomialTC}.
This intuition was confirmed by Ito \textit{et al.} \cite{Ito2025HardLabelExtraction}, who showed that special-case neurons require an exponential number of queries to extract as network depth increases.
Consequently, in practice, the assumptions required to extract DNNs in hard-label settings within polynomial time do not hold \cite{Ito2025HardLabelExtraction}.

In parallel, several fidelity-based model extraction methods have been proposed using physical attacks.
In \cite{Rakin2021DeepStealAM, Hector2023FaultIA}, fault injection is used to recover a subset of the weight bits, while constraint learning is applied to infer the remaining parameters.
Unlike the previously mentioned methods, these approaches are not restricted to fully connected DNNs and have successfully extracted complex architectures, such as ResNet-34 or VGG-11 \cite{Rakin2021DeepStealAM}.
However, these methods are either limited to DRAM-based inference platforms, such as Machine-Learning-as-a-Service (MLaaS) systems \cite{Rakin2021DeepStealAM}, or require access to an open device over which the attacker has full control, \textit{i.e.} a white-box setting—to locate the memory regions storing the weights \cite{Hector2023FaultIA}.
To the best of our knowledge, there is currently no method capable of extracting complex architectures, \textit{i.e.} not restricted to fully connected layers, without requiring access to confidence scores or an open device.

\paragraph{Contributions.} 
To overcome these issues, we present the first framework that combines side-channel attacks with cryptanalytic-based extraction of DNNs in hard-label settings.
To do so, we introduce a new generic methodology that subdivides the extraction problem into a divide-and-conquer strategy.
\corr{This approach offers higher precision than previous frameworks for weight extraction, improving the extraction of the targeted model while operating in a more restricted context.}
Furthermore, since our methodology is based on a different paradigm than previous ones, it is not impacted by the previously mentioned limitations\corr{, \textit{i.e.}, the extraction of special-case neurons, restrictions on layer types, and, most importantly, hard-label settings}.

We summarize the major contributions of our work as follows:
\begin{itemize}
    \item We highlight the limitations \corr{of extracting non-fully connected layers when state-of-the-art cryptanalytic-based methods are used in hard-label settings.}

    \item We introduce a new end-to-end methodology based on a \corr{side-channel oracle that is} more robust to special-case neurons and non-fully connected layers, thus \corr{enabling the extraction of complex DNNs.}
    
    \item \corr{To assess the suitability of the side-channel oracle, we apply a clustering-based side-channel attack to a state-of-the-art implementation of a constant-time ReLU activation function \cite{Maji2021LeakyNR}.}
    
    \item Finally, we validate our methodology by extracting all architectures targeted by state-of-the-art frameworks using $32$- and $64$-bit data, and by targeting two new architectures: an MLP with more than $1.7$ million parameters and a shortened version of MobileNetv1 embedded on an STM32F767ZI using the X-Cube-AI framework. This leads to the first cryptanalytic-based extraction of depthwise separable convolution, batch normalization, and pooling layers.    
\end{itemize}

We provide an implementation of our code at \url{https://github.com/bcoqueret/Side_channel_cryptanalytic_extraction_of_DNN}.

\paragraph{Threat Model.}\label{subsec:extraction_32bits_threat_model}
\corr{In the remainder of this work, we consider an attacker whose objective is to extract the weights of a DNN deployed on an embedded device, such as an MCU, in a hard-label setting.
Following prior work \cite{Jagielski2019HighAA, Rolnick2019ReverseengineeringDR, Carlini2020CryptanalyticEO}, we restrict our analysis to Deep ReLU networks and assume that the architecture of the target model is known to the attacker.
We do not, however, restrict the DNN to fully connected layers.}

\corr{As we consider a side-channel setting, we assume that the attacker has physical access to the device executing the target DNN, enabling the acquisition of side-channel measurements (\textit{e.g.}, electromagnetic (EM) traces in our case) during inference.
In contrast to existing works (\textit{e.g.}, \cite{Zhang2023DeepLearningME, Hector2023FaultIA}), we consider an attacker using non-profiled attacks based on unsupervised learning methods.
Consequently, the attacker has no \textit{a priori} knowledge of the model parameters and does not have access to the training or test datasets used to train the target DNN.}

\corr{We summarize in \autoref{tab:cmp_sota_all} the state-of-the-art (SOTA) frameworks performing \textit{fidelity-based model extraction} using cryptanalytic methods, fault injection, or side-channel attacks, as well as our contribution.}

\begin{table*}[ht]
\caption{Overview of the state-of-the-art of fidelity-based model extraction attacks.}
\label{tab:cmp_sota_all}
\fontsize{10pt}{10pt}\selectfont
\centering
\resizebox{\textwidth}{!}{
\begin{threeparttable}
  \begin{tabular}{|p{0.11\textwidth}| p{0.14\textwidth} | p{0.15\textwidth} | p{0.1\textwidth} | p{0.2\textwidth}| p{0.1\textwidth}| p{0.09\textwidth}| p{0.2\textwidth}|}
\hline
  \textbf{Approach} & \textbf{Attack type} & \textbf{Full extraction (weight + bias)} & \textbf{Hard-label setting\tnote{1}} & \textbf{Applicable to non-fully connected DNNs} & \textbf{Random queries\tnote{2}} & \textbf{DNN's datatype tested} & \textbf{Targeted architecture (Most complex)}\\
\hline
\cite{Rolnick2019ReverseengineeringDR} & Cryptanalytic & \cmark & \xmark & \xmark  & \cmark & 64-bit & MLP 10-20-20-1 \\
\cite{Carlini2020CryptanalyticEO} & Cryptanalytic & \cmark & \xmark & \xmark  & \cmark & 64-bit & MLP 40-20-10-10-1 \\
\cite{Chen2024HardLabelCE} & Cryptanalytic & \cmark & \cmark & \xmark  & \cmark & 64-bit & MLP 1024-2-2-1 \\
\cite{Carlini2024PolynomialTC} & Cryptanalytic & \cmark & \cmark & \xmark  & \cmark & 64-bit & MLP 3072-256$\times$3-64-10 \\
\hline
\cite{Batina2019CSINR} & Side-channel & \xmark & \cmark & \cmark  & \cmark & 32-bit & MLP 784-200$\times$4-10 \\
\cite{Zhang2023DeepLearningME} & Side-channel & \cmark & \xmark & \xmark  & \cmark & 64-bit & LeNet5\\
\hline
\cite{Rakin2021DeepStealAM} & Fault Injection & \cmark & \cmark & \cmark  & \xmark & 8-bit & ResNet34 and VGG11\\
\cite{Hector2023FaultIA} & Fault Injection & \cmark & \xmark & \cmark  & \xmark & 8-bit & CNN 3$\times$(Conv + Pooling + ReLU)-Linear\\
\hline
\textbf{This work} & Cryptanlytic and side-channel & \cmark & \cmark & \cmark  & \cmark & 32- and 64-bit& Shortened MobileNetv1 MLP 3072-256$\times$3-64-10\\
\hline
\end{tabular}
\begin{tablenotes}
    \item[1] Attacks which are \textbf{not} in hard-label settings suppose access to the confidence scores.
    \item[2] Contributions which are \textbf{not} using random queries use a subset of the training or a testing dataset to perform part of their attacks.
\end{tablenotes}
\end{threeparttable}}
\end{table*}
\section{Background}
\label{Background}

\subsection{Notations}
\label{subsec:notations}
Let calligraphic letters $\mathcal{X}$ denote sets, the corresponding capital letters $X$ (resp. bold capital letters) denote random variables (resp. random vectors $\textbf{T}$), and the lowercase $x$ (resp. $\textbf{t}$) denote their realizations.
We will use $\textbf{T}_i$ to describe the $i$-th element of a vector $\textbf{T}$.
Throughout this paper, the function modeled by a DNN is denoted as $f:\mathcal{X} \rightarrow |\mathcal{Y}|$, which characterizes its ability to classify a data $X \in \mathcal{X}$, \textit{e.g.}, an image, over a set of $|\mathcal{Y}|$ classes.
A DNN characterized by the function $f$ and the set of weights $\mathcal{\theta}$ is denoted $f_{\theta}$, and we denote $f_{\hat{\theta}}$ the model extracted from $f_{\theta}$.
The probability of observing an event $X$ is denoted by $\texttt{Pr}[X]$.
Finally, we denote $\mathcal{D}_{\mathcal{X}}$, the distribution over the set of data $\mathcal{X}$. 

\subsection{Fidelity-based model extraction}
\label{subsec:fidelity-extraction}

Model extraction attacks target the confidentiality of a deployed model by attempting to obtain a copy of the victim’s model.
The most common objective is to use the \corr{copied model} to benefit from the target’s capabilities, \textit{e.g.}, commercial value or performance on specific tasks, without having to train a model.
Another possibility is to use the \corr{copied} model to gain information about the target and mount higher-level attacks against it, such as the generation of adversarial examples in a white-box setting \cite{Szegedy2013IntriguingPO, Carlini2016TowardsET} or membership inference attacks \cite{Fredrikson2015ModelIA}.
This diversity of attack scenarios leads to several adversarial goals in model extraction.
Following the notation introduced in \cite{Jagielski2019HighAA}, we consider two main goals, namely \textit{task accuracy} and \textit{fidelity}.

The first goal aims to extract a model $f_{\hat{\theta}}$ such that, for the true task distribution $\mathcal{D}_{\mathcal{X} \times \mathcal{Y}}$ over the input space $\mathcal{X}$ and label space $\mathcal{Y}$, $f_{\hat{\theta}}$ maximizes $\texttt{Pr}_{X, Y}[\texttt{argmax}(f_{\hat{\theta}}(X)) = Y]$.
In practice, this can be achieved through \textit{learning-based} methods and optimization.
However, this approach introduces variability in the optimization process, which can lead to very different solutions, \textit{i.e.} models, due to convergence to different local minima.
As a result, this method offers no guarantee from a \textit{fidelity} perspective and is therefore suited for the objective of \textit{task accuracy}.

The second adversarial goal consists of maximizing the similarity between the \corr{copied} and the original models in their predictions.
The strongest form of this attack is \textit{functional equivalence}, which corresponds to obtaining an exact copy of the targeted DNN.
In this paper, we focus exclusively on \textit{fidelity-based model extraction}.
To evaluate the success of such extraction, we use \corr{the notion of ($\epsilon,\delta$)-functional equivalence}.
\begin{definition}[($\epsilon,\delta$)-functional equivalence \cite{Carlini2020CryptanalyticEO}]\label{def:relaxed-functionnal-eq}
Two models $f_{\theta}$ and $g_{\gamma}$ are ($\epsilon,\delta$)-functionally equivalent on $\mathcal{X}$ if:
\begin{equation*}
\begin{split}
    \texttt{Pr}_{X\in\mathcal{X}}[|f_{\theta}(X) - g_{\gamma}(X)| \leq \epsilon ] \geq 1 - \delta. \\
\end{split}
\end{equation*}
\end{definition}

The advantages of \textit{fidelity-based model extraction} have motivated extensive research in this area.
In particular, side-channel attacks \cite{Batina2019CSINR, Joud2022API, Zhang2023DeepLearningME, horvath2025barracuda} and fault-based attacks \cite{Rakin2021DeepStealAM, Hector2023FaultIA} have been used to target either all weights of a DNN individually or partial information about them.

\subsection{Side-channel attacks} 
\label{subsec:background_side_channel}
\corr{Historically, side-channel attacks (SCAs) are a class of cryptographic attacks in which an adversary exploits implementation-level vulnerabilities of real-world cryptosystems by analyzing physical leakages, such as power consumption or electromagnetic (EM) emissions.
During cryptographic execution, side-channel traces correlated with key-dependent intermediate variables are recorded.
Key discrimination is then achieved by applying statistical distinguishers to a large number of traces across all key hypotheses.
In \cite{Batina2019CSINR}, Batina \textit{et al.} were among the first to adapt these techniques to DNNs, targeting both architecture and parameters using Pearson correlation in a Correlation Electromagnetic Analysis (CEMA) combined with a Simple Power Analysis (SPA).
They specifically target the Hamming weight of the accumulated sum between a neuron’s weight vector and its input vector on ARM Cortex-M3 and ATmega328P devices.
However, limitations related to the ReLU function, error propagation, and bias extraction were identified in \cite{Joud2022API}.
Since then, CEMA has been extended to extract DNN weights on microcontrollers \cite{Joud2022API} and GPUs \cite{horvath2025barracuda}.
Several other side-channel techniques originally developed for cryptographic systems have also been adapted to DNNs.
In particular, deep learning–based side-channel attacks \cite{inproceedings, Cagli2017ConvolutionalNN} have been successfully applied to extract DNN hyperparameters.
Gao \textit{et al.} \cite{Gao2023DeepTheftSD} trained meta-models on physical traces to recover model architectures with high precision.
Earlier, Maia \textit{et al.} \cite{Maia2021CanOH} reconstructed ResNet and VGG architectures running on GPUs using physical traces.
All these attacks rely on supervised learning to construct the distinguisher, which requires access to an open device providing ground-truth labels.
This assumption is restrictive and motivates the exploration of more permissive threat models based on unsupervised learning, as specified in our threat model (see \autoref{introduction}).}

These approaches thus exhibit limitations in terms of attack complexity, supported architectures, or threat model assumptions (\corr{see \autoref{tab:cmp_sota_all}}).
On the other hand, cryptanalytic-based attacks have proven effective against shallow networks \cite{Milli2018ModelRF, Jagielski2019HighAA} and were later extended to deeper architectures \cite{Rolnick2019ReverseengineeringDR, Carlini2020CryptanalyticEO}.
These methods are briefly introduced in the following section.

\subsection{Cryptanalytic extraction methodology}
\label{subsec:Cryptanalytic-extraction}

\corr{Both methods in \cite{Rolnick2019ReverseengineeringDR, Carlini2020CryptanalyticEO} extract DNN weights by iteratively targeting each neuron and monitoring its state in order to identify its critical points and the associated neuron-induced hyperplane.
These concepts are defined as follows.}
\begin{definition}[Critical point and state of a neuron \cite{Carlini2020CryptanalyticEO}]\label{def:critical-pts-state}
Given an input $X \in \mathcal{X}$, let $V(\eta; X)$ be the function characterizing the input of the neuron $\eta$ before applying the activation function.
Then $X$ is said to be a critical point to the neuron $\eta$ if $V(\eta; X) = 0$.
If $V(\eta; X) > 0$ (resp. $V(\eta; X) < 0$) then $\eta$ is said to be active (resp. inactive).
\end{definition}
\begin{definition}[Neuron-induced Hyperplane]\label{def:hyperplane}
The hyperplane of a neuron $\eta$ corresponds to the set $\mathcal{H_{\eta}} \subseteq \mathcal{X}$ where $\forall X \in \mathcal{H_{\eta}}, V(\eta; X) = 0$.
\end{definition}

\begin{figure}[htbp]
    \centering
    \includegraphics[width=0.4\textwidth]{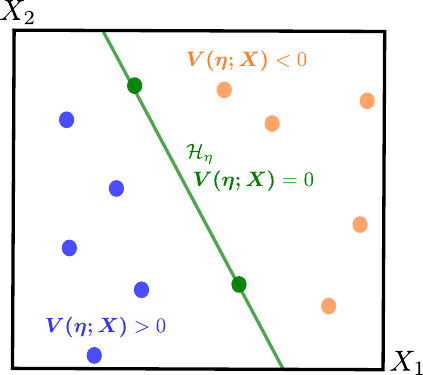}
    \caption{Neuron-induced hyperplane.}
    \label{fig:schema_background}
\end{figure}

An example of critical points, neuron states, and the neuron-induced hyperplane is shown in \autoref{fig:schema_background}.
\corr{Identifying critical points is of high importance, as it allows the determination of the neuron-induced hyperplane equation.}
This equation can then be used to recover the neuron’s weights with high precision:
\begin{equation*}
\tag{1}
\label{eq:critical_pts}
\begin{split}
    \sum_{i=0}^{n} \boldsymbol{\theta}_{\eta,i}\boldsymbol{X}_i + \beta_{\eta} = 0, \boldsymbol{X} \in \mathbb{R}^{n+1},\\
\end{split}
\end{equation*}
with $\boldsymbol{\theta_{\eta}}$ (resp. $\beta_{\eta}$) the weight vector (resp. the bias) associated with the neuron $\eta$.
It is important to note that for a neuron $\eta$ with $n + 1$ parameters, the induced hyperplane is an $n$-dimensional piecewise-linear surface \cite{Hanin2019DeepRN}.
\corr{Thus, from \autoref{eq:critical_pts}, we obtain,}
\begin{equation*}
\tag{2}
\label{eq:critical_pts2}
\begin{split}
    \boldsymbol{X}_0 &= \sum_{i=1}^{n} \frac{-\boldsymbol{\theta}_{\eta,i}}{\boldsymbol{\theta}_{\eta,0}}X_i - \frac{\beta_{\eta}}{\boldsymbol{\theta}_{\eta,0}}, \boldsymbol{\theta}_{\eta,0} \neq 0.
\end{split}
\end{equation*}

After extracting a critical point, the neuron’s weights can be recovered up to a \textit{scaling factor} by sampling along the standard basis vectors of $\mathcal{X}$.
This process is referred to as \textit{signature search} in \cite{Carlini2020CryptanalyticEO}.
\begin{definition}[Neuron signature \cite{Shamir2023PolynomialTC}]\label{def:signature}
    The signature of a neuron $\eta$ corresponds to a vector $\boldsymbol{S_{\eta}} \in \mathbb{R}^{n+1}$, for which there exists a scalar $\alpha \in \mathbb{R}$ such that $ \boldsymbol{\theta_{\eta}} = \alpha\cdot\boldsymbol{S_{\eta}}$.
\end{definition}
In \autoref{eq:critical_pts2}, the vector $\boldsymbol{S_{\eta}} = (1, \frac{\boldsymbol{\theta}_{\eta,1}}{\boldsymbol{\theta}_{\eta,0}}, \ldots, \frac{\boldsymbol{\theta}_{\eta,n}}{\boldsymbol{\theta}_{\eta,0}})$ is a signature of the neuron $\eta$ and $\boldsymbol{\theta}_{\eta,0}$ is the scaling factor.
Since the ReLU function is equivariant under positive scaling, \textit{i.e.}, $\forall x \in \mathbb{R}$ and $\forall c \in \mathbb{R}^{+}$, $\texttt{ReLU}(c \cdot x) = c \cdot \texttt{ReLU}(x)$, extracting neuron signatures rather than exact weight vectors allows faithful reconstruction.
As long as the sign of the ReLU input is preserved, activation values can be re-scaled in subsequent layers.
Thus, the precise value of the scaling factor is irrelevant; only its sign matters.
Consequently, this approach recovers a functionally equivalent model rather than an exact copy.\\

Under these assumptions, model extraction proceeds iteratively for each neuron in two steps:
\begin{enumerate}
  \item Extraction of the neuron's signature.
  \item Extraction of the sign of the scaling factor to ensure that the neuron's state remains the same.
\end{enumerate}

\corr{These steps are detailed in \autoref{subsec:history_cryptanalytic_extract}.}
\corr{While these methods have successfully enabled complete extraction of Deep-ReLU networks, significant challenges remain, such as the full extraction of non–fully connected DNNs in hard-label settings (see \autoref{tab:cmp_sota_all}).
These limitations motivate the development of a new methodology that combines cryptanalytic extraction with side-channel analysis.}

\section{\corr{Motivations and extension using side-channel attacks}} \label{sec:extraction_32bits_motivations}

Although the methods presented in the previous section demonstrate strong results, \textit{fidelity-based model extraction} remains a difficult task.
\corr{First, in \autoref{subsec:history_cryptanalytic_extract}, we provide a recap of the different steps required to perform cryptanalytic-based extraction attacks.}
\corr{Then, in \autoref{subsec:natural_limit}, we present inherent difficulties that have not been addressed in these attacks, \textit{e.g.} restrictions on the targeted architecture, the high precision required to estimate gradients, and special cases of neurons.
Finally, in \autoref{subsec:divide_conquer_methodo}, we describe a divide-and-conquer methodology using a side-channel oracle to tackle the problems identified in \autoref{subsec:natural_limit}.}

\subsection{\corr{Historical methodology for cryptanalytic-based extraction}}
\label{subsec:history_cryptanalytic_extract}
\corr{As introduced in \autoref{subsec:Cryptanalytic-extraction}, multiple steps are required to conduct a cryptanalytic-based extraction attack.
First, an attacker must retrieve each neuron’s signature by identifying a set of critical points.
This step requires determining the neuron’s state and, consequently, extracting the corresponding signature.
Then, given a set of signatures, the attacker must identify the sign of the scaling factor of each neuron in order to correctly perform fidelity-based model extraction.
This section provides an overview of these steps and highlights the most recent state-of-the-art contributions extending the seminal work.}

\paragraph{Extraction of the neuron's signature.} 
By considering the targeted model $f_{\boldsymbol{\theta}}$ as an oracle, the search for a neuron’s signature can be performed directly using its input–output pairs.
Indeed, if the targeted model is composed of piecewise-linear activation functions, then $f_{\boldsymbol{\theta}}$ is also piecewise linear, and discontinuities can be observed in the gradient of $f_{\boldsymbol{\theta}}$.
These discontinuities correspond to changes in the activation state of neurons within the DNN.
More specifically, the discontinuity in the gradient of the ReLU function is localized at $0$, which implies that by monitoring the gradient of the targeted model, critical points for each neuron can be identified.
However, performing signature extraction using these discontinuities remains challenging and requires strong assumptions.
First, the target model’s output must expose either confidence scores or logits, enabling gradient estimation through finite differences.
Second, since it is impossible to determine which neuron corresponds to a given critical point, a large number of critical points must be sampled to ensure that each neuron is represented.
Third, in deeper layers, on average, half of the neurons are deactivated \cite{Carlini2020CryptanalyticEO}, which implies that many entries of a critical point are zero and therefore prevent recovery of the associated weights.
Carlini \textit{et al.} \cite{Carlini2020CryptanalyticEO} address this issue by recombining signatures obtained from different critical points.
However, this requires ensuring that the signatures correspond to the same neuron and layer.
To mitigate this, they introduce a consistency check based on the assumption that signatures from different neurons or layers lead to inconsistent results.
Finally, the model output must be returned with high numerical precision to ensure accurate gradient estimation.
In \cite{Jagielski2019HighAA, Carlini2020CryptanalyticEO}, the authors successfully extract ReLU-based networks operating with double-precision (\textit{i.e.}, 64-bit floating-point) arithmetic.
\corr{While neuron signatures can be revealed through gradient discontinuities, the sign of the scaling factor is not directly accessible to the attacker.
Consequently, an additional extraction step is required.}

\paragraph{Extraction of the neuron's sign.} 
Carlini \textit{et al.} \cite{Carlini2020CryptanalyticEO} extract the sign of the neurons by exhaustively testing the $2^m$ possible sign configurations, where $m$ denotes the number of neurons in the layer.
As this approach quickly becomes impractical for large DNNs, Canales-Martínez \textit{et al.} \cite{Shamir2023PolynomialTC} propose an alternative method based on crafting inputs that selectively activate specific neurons to infer their sign.
Their approach successfully extracts the signs of neurons in a model with eight hidden layers, each containing $256$ neurons.
\corr{This process is applied sequentially (\textit{i.e.}, for each neuron in each layer), starting from the first layer and proceeding toward the last.}
As a result, this attack operates in a gray-box setting, assuming knowledge of the architecture and access to confidence scores.

\paragraph{\corr{New directions in the SOTA.}} 
Chen \textit{et al.} \cite{Chen2024HardLabelCE} were the first to propose a cryptanalytic extraction method for DNNs operating in hard-label settings, targeting binary classifiers.
Their approach relies on \textit{decision boundary points}, \textit{i.e.}, inputs located on the boundary between two labels, and was demonstrated on DNNs with two hidden layers.
However, this method exhibits exponential time complexity \cite{Carlini2024PolynomialTC} and is limited to binary classification, motivating the search for alternative approaches.
Carlini \textit{et al.} \cite{Carlini2024PolynomialTC} introduce such an alternative by leveraging decision boundary points\footnote{The authors in \cite{Carlini2024PolynomialTC} refer to these points as transition points. Out of clarity we choose to use the notation decision point in this work.} to identify critical points.
Due to the piecewise-linearity of ReLU networks, the input space is partitioned into linear regions, referred to as \textit{cells}.
In prior methods, boundaries between cells, corresponding to neuron-induced hyperplanes, were identified through gradient estimation, which is not feasible in hard-label settings.
The key insight of \cite{Carlini2024PolynomialTC} is that decision boundaries must be flat hyperplanes within a given cell.
Thus, they identify \textit{dual points}, defined as points that lie both on a decision boundary and on a cell boundary, by analyzing the geometry of the decision boundary.
An illustration of such points is shown in \autoref{fig:dual_point_no_pooling}.
Since these points are also critical, they can be used to extract neuron signatures (see \autoref{def:signature}).
\corr{However, this strategy presents several limitations, which are discussed in \autoref{subsec:natural_limit}.}

\corr{Model extraction in hard-label settings is not the only recent direction explored in the SOTA.
Indeed, since the last layer of a DNN typically does not use a ReLU activation function, its extraction requires a different strategy.}
In soft-label settings, attackers can exploit confidence scores as labels for supervised learning \cite{Jagielski2019HighAA}.
In contrast, hard labels do not provide sufficient information to guarantee functional equivalence, and extraction of the last layer was not addressed in \cite{Carlini2024PolynomialTC}.
Recently, Canales \textit{et al.} \cite{Canales_extraction_last_layer} proposed a first methodology to address this issue.
Their approach uses points on the decision boundary to form a system of equations that enables extraction of the final layer’s parameters.
However, such systems are not always full rank.
To address this, the authors fix a subset of variables, allowing recovery of an equivalent model.
Although this method does not guarantee exact parameter recovery, the extracted model achieves perfect fidelity with respect to the target \cite{Canales_extraction_last_layer}.

\corr{While these developments improve the practicality of cryptanalytic-based extraction, several open limitations remain, which are discussed in the following section.}

\subsection{Natural difficulties in cryptanalytical extraction of DNNs}\label{subsec:natural_limit}
Due to the complexity of cryptanalytical extraction, several limitations persist.
The most direct limitations of the methodologies presented in \cite{Jagielski2019HighAA, Carlini2020CryptanalyticEO} are \corr{those related to the threat model}.
\corr{On the first hand, gradient estimation requires access to confidence scores or logits with high numerical precision (\textit{i.e.}, 64-bit floating-point values).
On the other hand, no prior work has successfully demonstrated extraction of DNNs that include non-fully connected layers, such as max or average pooling layers.
While the first limitation appears natural, the second requires further explanation.}

\paragraph{Limitation to extract non-fully connected layers.}
As mentioned in \autoref{subsec:history_cryptanalytic_extract}, the most promising methodology in hard-label settings was introduced in \cite{Carlini2024PolynomialTC}.
In this work, Carlini \textit{et al.} propose using the decision boundary to identify dual points, \textit{i.e.} points that lie both on the decision boundary and on a neuron-induced hyperplane, to extract trained parameters, namely weights and biases.
They successfully demonstrate extraction of all hidden layers of a four-layer DNN trained on CIFAR-10 using only hard labels.
However, their evaluation is limited to fully connected networks.
We argue that this methodology is not directly applicable to more complex architectures, especially those including non-fully connected layers such as max-pooling layers.

\begin{figure}[ht]
\centering
    \begin{subfigure}{.5\linewidth}
        \includegraphics[width=\linewidth]{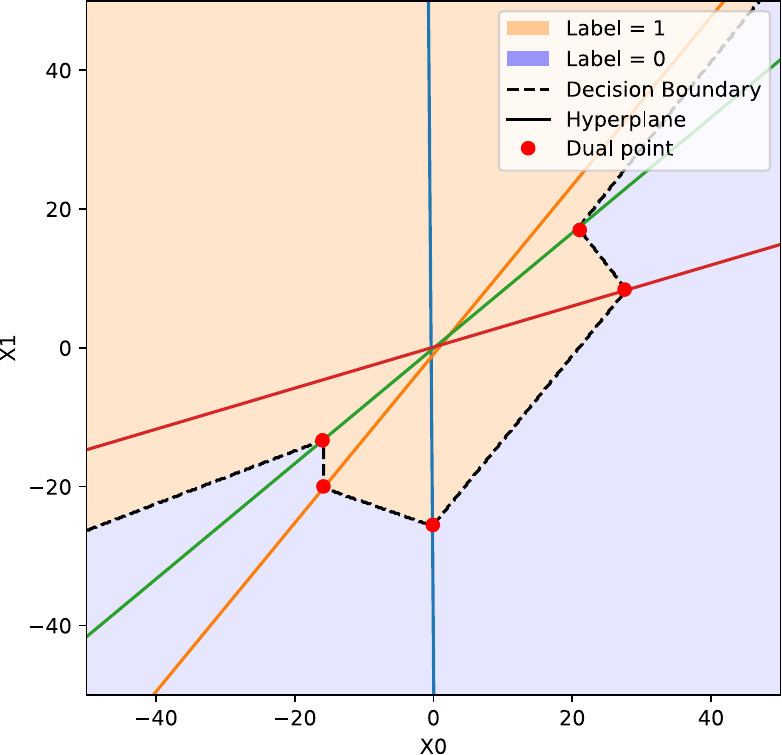}
        \caption{Fully connected network.}
        \label{fig:dual_point_no_pooling}
    \end{subfigure}%
    \hfill
    \begin{subfigure}{.5\linewidth}
        \includegraphics[width=\linewidth]{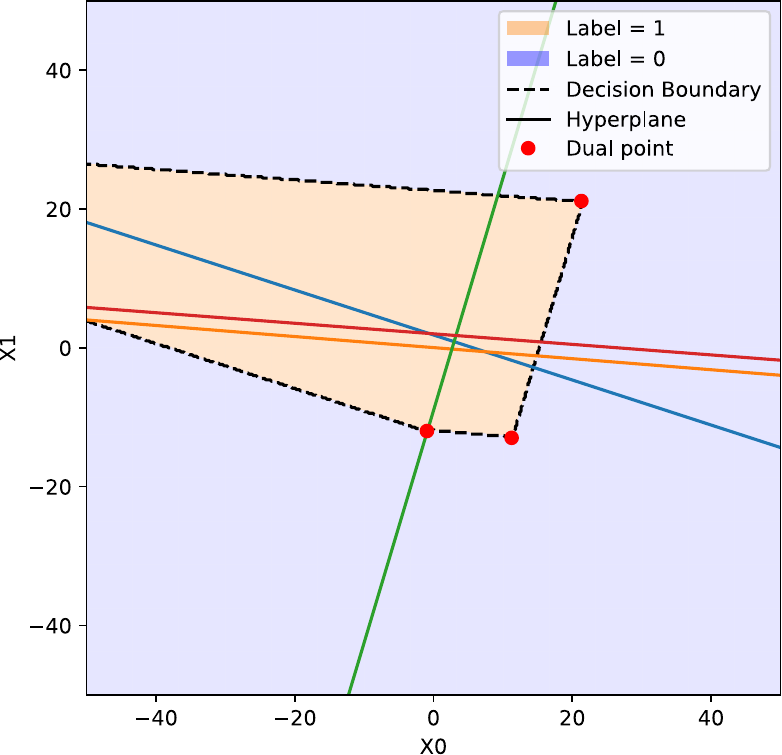}
        \caption{DNN with a max pooling layer.}
        \label{fig:dual_point_with_pooling}
    \end{subfigure}
    \caption{Illustration of the concept of dual points as introduced in \cite{Carlini2024PolynomialTC} on a MLP with one hidden layer of four neurons each associated with a different hyperplane represented by the colored lines.}
    \label{fig:schema_method_dual}
\end{figure}

To illustrate this limitation, we train a simple linear neural network with one hidden layer on a binary classification task.
In \autoref{fig:dual_point_no_pooling}, we observe that the shape of the decision boundary changes only when crossing a neuron-induced hyperplane, which is the core assumption of the method in \cite{Carlini2024PolynomialTC}.
In this case, the identified points effectively correspond to dual points, and the hyperplane can be recovered in hard-label settings.
However, when training the same network with the addition of a max-pooling layer, we obtain comparable classification performance, yet observe in \autoref{fig:dual_point_with_pooling} that the decision boundary no longer correlates with the neuron-induced hyperplanes.
This discrepancy leads to incorrect identification of dual points and, consequently, incorrect estimation of weights.
Therefore, in such configurations, the approach proposed in \cite{Carlini2024PolynomialTC} is no longer applicable, and cryptanalytical model extraction fails.
This behavior can be explained by the fact that while a change of linear cell may occur, the max-pooling operation can suppress this change if the corresponding activation is not selected by the pooling layer.
As a result, the change is not visible in the decision boundary.
\corr{Hence, the attacker cannot correctly identify dual points, leading to poor extraction results.
While historical approaches do not scale well to non-fully connected layers, additional limitations also arise in more favorable settings, \textit{i.e.} when targeting fully connected layers.}

\paragraph{Limitation to extract fully connected layers.}
\corr{Even when extracting fully connected layers, which represents an ideal scenario for the attacker, cryptanalytic extraction using the methodology described in \autoref{subsec:history_cryptanalytic_extract} remains challenging for several reasons.
We highlight three issues that are addressed in \autoref{subsec:divide_conquer_methodo}.}

A first issue, reported in \cite{Shamir2023PolynomialTC}, concerns special cases of neurons that almost never change state.
Two types of such neurons are identified:
\begin{itemize}
    \item[$\bullet$] \textbf{\textit{Always-Off}.} These correspond to \textit{dead} neurons whose parameters cause their pre-activation value to be negative for most inputs.
    In ReLU-based DNNs, their output is therefore always zero.

    \item[$\bullet$] \textbf{\textit{Always-On}.} These neurons exhibit the opposite behavior, with pre-activation values that are always positive.
    They therefore behave as linear transformations.
\end{itemize}
For these neurons, the number of queries required to identify critical points can exceed that of typical neurons by an order of magnitude (see \autoref{subsec:extraction_32bits_special_cases}).
Despite their practical importance, no prior work provides a systematic solution for handling these cases.

A second issue is that input–output pairs do not allow direct identification of the neuron associated with a given critical point.
Consequently, attackers cannot verify whether all neurons have been successfully extracted (\textit{i.e.} all neurons have critical points).
Carlini \textit{et al.} \cite{Carlini2020CryptanalyticEO} address this by sampling a large number of critical points relative to the number of neurons, ensuring coverage with high probability.
However, this leads to redundant queries and significantly increases attack complexity, particularly for deep architectures.

Finally, the absence of direct mapping of the critical points to the targeted neuron leads to another problem brought to light in \cite{Liu2025NavigatingTD}.
More specifically, their concern is that the consistency check used by Carlini \textit{et al.} in \cite{Carlini2020CryptanalyticEO} to aggregate the incomplete signature of a given neuron is not correct.
\corr{In deeper layers, ReLU deactivations often produce incomplete signatures because some input components are suppressed. 
Since extracting a neuron requires a complete signature, Carlini et al. \cite{Carlini2020CryptanalyticEO} propose aggregating multiple partial signatures to recover missing entries. 
Their method merges signatures by verifying that their shared entries are proportional, increasing the aggregated signature size with each merge. 
They argue that aggregated signatures of size greater than two must originate from the targeted layer, as deeper-layer signatures are unlikely to satisfy the merge condition. 
However, Liu et al. show that signatures from deeper layers can also be merged when neurons share identical activation patterns across intermediate layers, causing the consistency check to fail and resulting in incorrect signature recovery.}
Several solutions to mitigate this issue have been introduced in \cite{Liu2025NavigatingTD}, which have led to the successful extraction of signatures from deeper layers.
More particularly, they are able to increase the SOTA by extracting $95\%$ of the weights in each layer of DNN with eight hidden layers on MNIST, while previous methodologies could not perform the extraction beyond the third layer \cite{Liu2025NavigatingTD}.
However, even with their method, targeting very deep architectures remains difficult.
Furthermore, while these improvements make the attacks more practical, they do not change the threat model and still consider full access to the confidence scores.\\

\corr{This section highlights the need for a new methodology for cryptanalytic-based model extraction.
In the following section, we introduce such a methodology by adopting a divide-and-conquer strategy.}



\subsection{\corr{Divide-and-conquer methodology}}\label{subsec:divide_conquer_methodo}
\begin{figure}[htbp]
    \begin{center}
       \includegraphics[width=.8\linewidth]{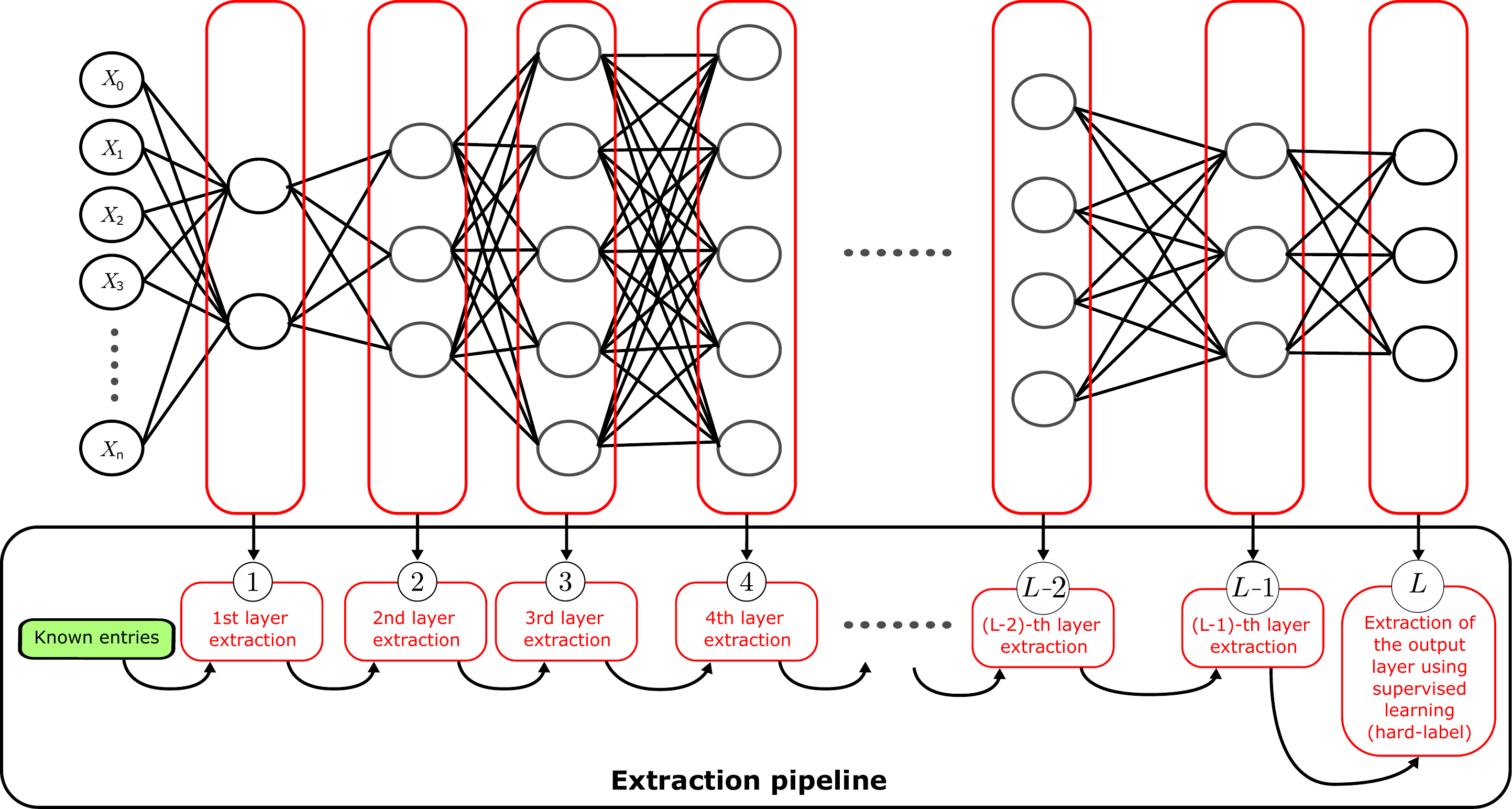}
    \end{center}
    \caption{High-level description of the divide-and-conquer approach when targeting a $L$ hidden layers DNN.} 
    \label{fig:highlevel_schema}
\end{figure}

\corr{This section provides a high-level overview of the proposed approach, while \autoref{sec:extraction_32bits_methodo} presents a detailed description of each step.
Our objective is to introduce an alternative fidelity-based model extraction methodology that is robust to special-case neurons and capable of extracting deep networks composed of non-fully connected layers in hard-label settings.
We argue that many of the existing limitations arise from targeting the model as a whole and do not apply to single-layer neural networks (NNs).
To facilitate cryptanalysis-based extraction attacks in hard-label settings, and thereby mitigate the limitations discussed in \autoref{subsec:history_cryptanalytic_extract} and \autoref{subsec:natural_limit}, we propose an approach based on a side-channel oracle $\mathsf{O}$ that decomposes the problem into the extraction of successive single-layer NNs.
Accordingly, a DNN with $L$ layers is decomposed into $L$ one-layer NNs, where the input to the $\ell$-th network is the output of the $(\ell-1)$-th network, for $\ell \in [1, L]$.
This divide-and-conquer strategy reduces the attack complexity by enabling the analysis of the DNN’s internal variables rather than treating the network as a monolithic system.
A high-level illustration of the extraction pipeline is shown in \autoref{fig:highlevel_schema}.
The remainder of this section focuses on the extraction of a single-layer NN; extracting a DNN with $L$ layers then requires repeating this process $L-1$ additional times.}

\corr{As specified in \autoref{subsec:history_cryptanalytic_extract}, cryptanalytic-based extraction requires two steps to extract one neuron of the $\ell$-th layer: one for extracting the neuron's signature and one for determining the sign of the scaling factor.
As a reminder, extracting a neuron's signature requires identifying critical points for each neuron of the $\ell$-th layer.
To perform this identification, we use a side-channel oracle $\mathsf{O}$, which returns a boolean value equal to \textit{True} if two neurons are associated with the same state for two given inputs.}
We justify this choice by the fact that activation functions of DNNs have already been targeted using distinguishers exploiting timing differences in \cite{Batina2019CSINR, Maji2021LeakyNR}, leading to the identification of the activation function and even the weights' mantissa.
While effective, both previously mentioned approaches rely on supervised learning and therefore require access to an open device to train the distinguisher.
In this work, we consider a less restrictive adversary who does not require such an open device to conduct the attack (see \autoref{tab:cmp_sota_all}).
Consequently, a supervised learning strategy is not applicable.
To address this limitation, we propose an alternative distinguisher based on an unsupervised \textit{k-means} clustering algorithm \cite{kmeanbook}.
The idea is to generate a sufficient number of EM traces using random inputs to build a dataset in which both classes, \textit{i.e.}, active and inactive neurons, are evenly represented.
We then apply the \textit{k-means} algorithm to identify two clusters corresponding to the two neuron states.
As mentioned previously (see \autoref{sec:extraction_32bits_methodo}), it is not initially necessary to determine which cluster corresponds to the active or inactive state, since this information is not required to identify a critical point.
These clusters are then used as references to infer the state of a neuron.
Finally, given the EM trace of a new neuron, we use the \textit{Euclidean} distance to the centroids of the two clusters as a distinguisher to determine the neuron's state.
Under the assumption that the implementation of the ReLU function is identical for all neurons across all layers, this distinguisher can be used as an oracle to perform the complete extraction.\\

\corr{An additional advantage of such a side-channel oracle is the direct mapping from a critical point to its corresponding neuron.
Indeed, as discussed in \autoref{subsec:natural_limit}, a key limitation of classical cryptanalytic methods is their inability to determine which neuron corresponds to a given critical point, resulting in the extraction of unnecessary critical points.
This limitation is not present in our methodology, as the oracle directly provides the state of the targeted neuron.
We are therefore able to localize which neuron corresponds to each identified critical point.
Furthermore, this direct mapping naturally eliminates the issues related to merging signatures described in \cite{Liu2025NavigatingTD}.}\\

\begin{figure}[htbp]
    \begin{center}
       \includegraphics[width=.97\linewidth]{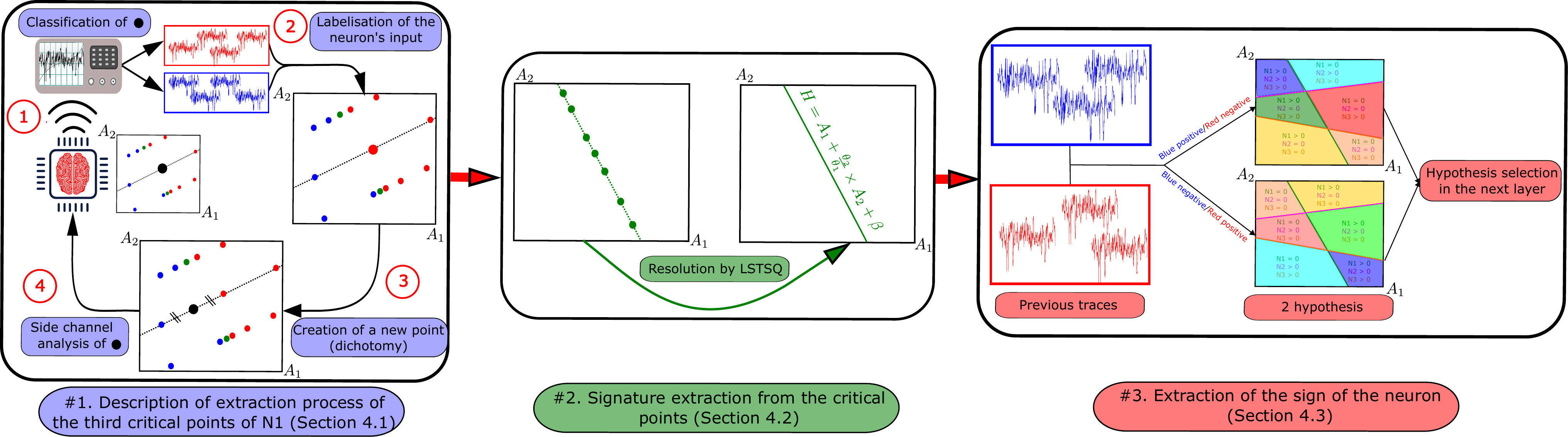}
    \end{center}
    \caption{Description of our complete methodology for extracting a neuron included into the $\ell$-th layer.} 
    \label{fig:global_schema_iterative}
\end{figure}

\corr{Given this high-level description of the divide-and-conquer strategy, we summarize the extraction of each neuron of the $\ell$-th layer into three stages. 
This strategy is illustrated in \autoref{fig:global_schema_iterative}:}
\begin{enumerate}

  \item[$\bullet$] During the first stage of our methodology, we use a side-channel-based oracle to perform a binary search to identify critical points, following \autoref{alg:extraction_32bits_algo_search}.
  \corr{In contrast to historical approaches (\textit{e.g.} \cite{Jagielski2019HighAA, Carlini2020CryptanalyticEO}), this contribution offers two main benefits: first, it does not require access to confidence scores or logits to identify critical points; second, the oracle can also be used to extract non-fully connected layers such as pooling layers.
  This stage is detailed in \autoref{subsec:extraction_32bits_split} and discussed in \autoref{subsec:discuss_methodo}.}
  
  \item[$\bullet$] The second stage consists of approximating the equation describing the hyperplane induced by the targeted neuron.
  The critical points identified in Stage~1 are used following the method introduced in \autoref{subsec:extraction_32bits_signature}.
  \corr{This stage also enables, for the first time, the extraction of special-case neurons (\textit{i.e.}, always-off, always-on, and input-off neurons).}
  
  \item[$\bullet$] \corr{In the third stage, we propose a new approach to extract the sign of each neuron by leveraging the side-channel oracle.
  Described in \autoref{subsec:extraction_32bits_sign}, this approach allows the extraction of the sign of each neuron in the $\ell$-th layer without requiring additional queries to the targeted DNN.}
  
\end{enumerate}

\corr{Once the three stages have been performed for the $(L-1)$ one-layer NNs, the final layer must be extracted.
As specified in \autoref{subsec:history_cryptanalytic_extract}, the extraction process for the last layer differs in hard-label settings.
While Canales \textit{et al.} \cite{Canales_extraction_last_layer} use points on the decision boundary to extract the final trainable parameters, we instead employ supervised learning on a dataset composed of the random inputs generated for the previous layers and their corresponding hard labels.
This dataset is used to train the last layer to behave similarly to that of the targeted model.
Although this methodology enables extraction of the final layer in hard-label settings, a discussion is provided in \autoref{subsec:extraction_32bits_last_layer} to evaluate the impact of this strategy.}
\section{Subdividing DNN to improve cryptanalytical extraction} \label{sec:extraction_32bits_methodo}

\corr{In this section, we detail the \corr{three-stage} methodology introduced in \autoref{subsec:divide_conquer_methodo}.}
\corr{In \autoref{subsec:extraction_32bits_split}, we describe how to use a \corr{side-channel} oracle function to identify critical points.} 
Then, in \autoref{subsec:extraction_32bits_signature}, we illustrate how this set of critical points can be used to extract the parameters of each layer and \corr{introduce a new strategy to handle special-case neurons.}
Finally, in \autoref{subsec:extraction_32bits_sign}, we present a method to extract the signs of all scaling factors within a layer using a single hypothesis.

\subsection{Extraction of critical points} \label{subsec:extraction_32bits_split}

In our methodology, we assume access to a \corr{side-channel} oracle $\mathsf{O}:\mathbb{N} \times \mathbb{N} \times \mathcal{X} \times \mathcal{X} \rightarrow \mathcal{B}$, which returns a boolean value indicating whether two inputs $\boldsymbol{X} \in \mathcal{X}$ and $\boldsymbol{X}^{\prime} \in \mathcal{X}$ are associated with the same state for two neurons $(\eta, \gamma) \in \mathbb{N}^2$, \textit{i.e.}, whether $V(\eta; \boldsymbol{X}) = V(\gamma; \boldsymbol{X}^{\prime})$ (see \autoref{def:critical-pts-state}). 
As such, the \corr{side-channel} oracle $\mathsf{O}$ returns \textsf{True} when the two inputs correspond to the same state, and \textsf{False} otherwise.
Since this function is often applied to different inputs for the same neuron $\eta$, we use the following simplified notation when no ambiguity arises: $\mathsf{O}(\eta, \boldsymbol{X}, \boldsymbol{X}^{\prime}) = \mathsf{O}(\eta, \eta, \boldsymbol{X}, \boldsymbol{X}^{\prime})$.
Moreover, throughout this work, we assume that $\mathcal{X} \subseteq \mathbb{R}^{n+1}$, with $n \in \mathbb{N}$.
This oracle naturally removes the need for soft labels in the search for critical points.

\begin{remark}
    \corr{Note that this oracle does not require knowledge of the true states $V(\eta; \boldsymbol{X})$ and $V(\eta; \boldsymbol{X}^{\prime})$ associated with the inputs $\boldsymbol{X}$ and $\boldsymbol{X}^{\prime}$.
    The only required information is whether $V(\eta; \boldsymbol{X})$ and $V(\eta; \boldsymbol{X}^{\prime})$ belong to the same state.}
\end{remark}

\corr{To identify critical points, an attacker samples random inputs $\boldsymbol{X}$ and $\boldsymbol{X}^{\prime}$ in $\mathcal{X}$ until she finds a tuple $(\boldsymbol{X}, \boldsymbol{X}^{\prime})$ such that $V(\eta; \boldsymbol{X}) \neq V(\eta; \boldsymbol{X}^{\prime})$.
She then updates the tuple $(\boldsymbol{X}, \boldsymbol{X}^{\prime})$ by performing a binary search using the side-channel oracle $\mathsf{O}$ until the distance between $\boldsymbol{X}$ and $\boldsymbol{X}^{\prime}$ is bounded by a parameter $\Delta$.}
A high-level description of this procedure is presented in \autoref{alg:extraction_32bits_algo_search}.
Since we perform a binary search, finding a true critical point for the targeted neuron $\eta$, \textit{i.e.}, a point $\boldsymbol{X}$ lying exactly on the hyperplane $\mathcal{H}_{\eta}$ (see \autoref{def:hyperplane}), is challenging due to numerical approximations.
Instead, we introduce a hyperparameter $\Delta$ to obtain a point located within a bounded distance from the hyperplane $\mathcal{H}_{\eta}$.

\begin{algorithm}
\caption{\corr{Binary search of a critical point using an oracle.}}\label{alg:extraction_32bits_algo_search}
\begin{algorithmic}
\Require a side-channel oracle $\mathsf{O}:\mathbb{N} \times \mathbb{N} \times \mathcal{X} \times \mathcal{X} \rightarrow \mathcal{B}$, the targeted neuron $\eta \in \mathbb{N}$ and a distance precision $\Delta \in \mathbb{R}$
\Ensure $\boldsymbol{X} \in \mathcal{H}_{\eta}$
\State $(\boldsymbol{X}, \boldsymbol{X}^{\prime}) \gets (\texttt{random}(\mathcal{X}), \texttt{random}(\mathcal{X}))$ \Comment{Generate a random tuple in $\mathcal{X}$}
\State $S \gets O(\eta, \boldsymbol{X}, \boldsymbol{X}^{\prime})$
\While{$S \text{ is }\textsf{True}$}
    \State $\boldsymbol{X}^{\prime} \gets \texttt{random}(\mathcal{X})$
    \State $S \gets O(\eta, \boldsymbol{X}, \boldsymbol{X}^{\prime})$
\EndWhile
\While{$||\boldsymbol{X} - \boldsymbol{X}^{\prime}||_2 > \Delta$}
\State $\boldsymbol{M} \gets \frac{\boldsymbol{X}+\boldsymbol{X}^{\prime}}{2}$
\State $S_M \gets O(\eta, \boldsymbol{X}, \boldsymbol{M})$
\If{$S_M \text{ is }\textsf{True}$} \Comment{Implies that $V(\eta; \boldsymbol{M}) = V(\eta; \boldsymbol{X})$}
    \State $\boldsymbol{X} \gets \boldsymbol{M} $
\Else \Comment{Implies that $V(\eta; \boldsymbol{M}) = V(\eta; \boldsymbol{X}^{\prime})$}
    \State $\boldsymbol{X}^{\prime} \gets \boldsymbol{M} $
\EndIf
\EndWhile \\
\Return $\boldsymbol{X}$
\end{algorithmic}
\end{algorithm}

Under the assumption that the \corr{side-channel} oracle operates perfectly, this algorithm is equivalent to a dichotomic search over a sorted list of binary values.
\corr{Since binary search is used to retrieve critical points, the complexity of this phase matches that of \cite{Jagielski2019HighAA}.
Specifically, it requires $\mathcal{O}\!\left(\log_2\!\left(\frac{L_0}{\Delta}\right)\right)$ queries to obtain a point within distance $\Delta$ of the hyperplane, where $L_0$ denotes the initial distance between the first pair $(\boldsymbol{X}, \boldsymbol{X}^{\prime})$ such that $\mathsf{O}(\eta, \boldsymbol{X}, \boldsymbol{X}^{\prime})$ is \textsf{False}.
With this side-channel oracle, it is therefore possible to search for all critical points within a layer.
Once a set of critical points has been extracted, the attacker must identify and extract the corresponding signatures to recover the hyperplane $\mathcal{H}$.}



\subsection{Improved signature extraction}
\label{subsec:extraction_32bits_signature}

Since each critical point provides an equation linking a known input\footnote{\corr{This holds since, to extract the $\ell$-th layer, it is assumed that the $(\ell-1)$-th layer has already been successfully extracted.}} to an unknown weight vector, it is straightforward to construct a system of equations to address this problem.

\paragraph{Generic method.} 
We infer the equation of the neuron-induced hyperplane (see \autoref{def:hyperplane}) by constructing a linear system based on \autoref{eq:critical_pts} using the critical points obtained from \autoref{alg:extraction_32bits_algo_search}. 
Such a system can then be used to extract the weights. 
Given a set of $n+1$ critical points $\{\boldsymbol{X}^{1}, \boldsymbol{X}^{2}, \ldots, \boldsymbol{X}^{n+1}\}$ such that $\boldsymbol{X}^{i} \in \mathbb{R}^{n+1}$, the following system of equations can be used to extract the $n$ weights and the bias of the targeted neuron $\eta$:
\begin{equation}
\label{eq:system_basique}
  \left\{
    \begin{aligned}
    \sum_{i=1}^{n} \frac{-\boldsymbol{\theta}_{\eta,i}}{\boldsymbol{\theta}_{\eta,0}}\boldsymbol{X}_i^{1} &- \frac{\beta_{\eta}}{\boldsymbol{\theta}_{\eta,0}} =\boldsymbol{X}_0^{1}  \\
      &\vdots \\
      \sum_{i=1}^{n} \frac{-\boldsymbol{\theta}_{\eta,i}}{\boldsymbol{\theta}_{\eta,0}}\boldsymbol{X}_i^{n+1} &- \frac{\beta_{\eta}}{\boldsymbol{\theta}_{\eta,0}} =\boldsymbol{X}_0^{n+1} \\
    \end{aligned}
  \right.
\end{equation}
Once a sufficient number of critical points has been sampled using \autoref{alg:extraction_32bits_algo_search} to construct a full-rank system, we recover the equation of the hyperplane by solving the corresponding LSTSQ problem.
We then extract the signature (see \autoref{def:signature}) of the targeted neuron by arbitrarily fixing the scaling factor $\theta_{\eta,0}$ to $\pm 1$.
The sign of this scaling factor determines the state of the neuron and is therefore a critical parameter to extract.
The extraction of the scaling factor sign is detailed in \autoref{subsec:extraction_32bits_sign}.

\begin{figure}[htbp]
    \begin{center}
       \includegraphics[width=0.95\linewidth]{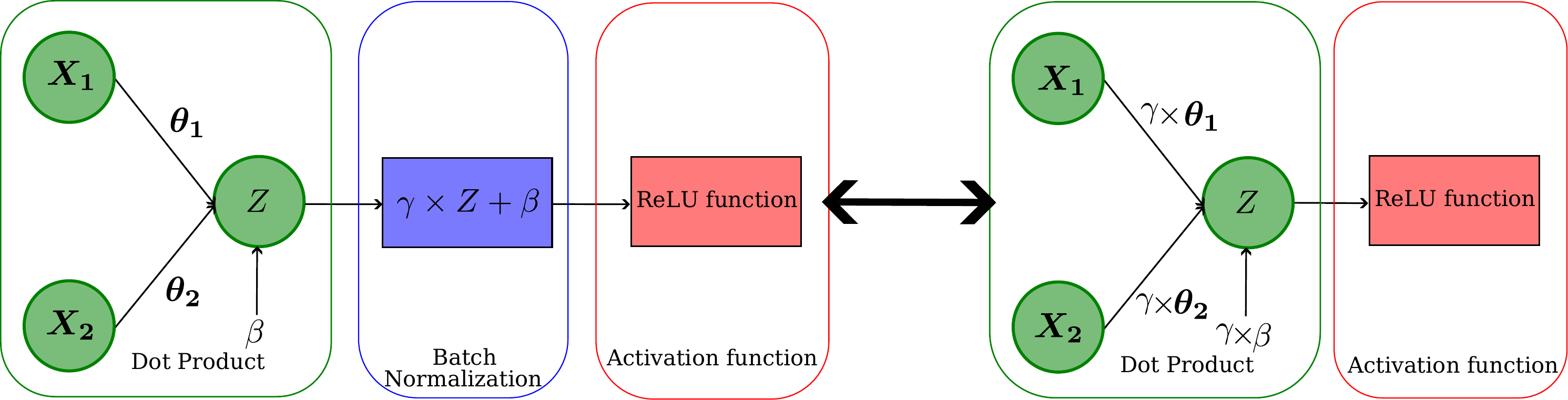}
    \end{center}
    \caption{Illustration of the merging process of a linear layer with batch normalization.}
    \label{fig:schema_fuzing_BN}
\end{figure}

\begin{remark}
    \corr{It is important to note that the presence of linear transformations between the matrix multiplication and the activation function, \textit{e.g.} batch normalization, does not impact the proposed method.}
    Indeed, all such linear operations can be merged into a single equivalent transformation, from which an aggregate set of weights can be extracted.
    This \corr{merging process} is illustrated in \autoref{fig:schema_fuzing_BN}. 
\end{remark}

This method extracts the signature of a targeted neuron using the critical points obtained from \autoref{alg:extraction_32bits_algo_search}, under the assumption that a full-rank system can be constructed.
This implies that the number of critical points, \textit{i.e.}, the number of equations, must be at least equal to the number of targeted parameters (weights and bias), and that each targeted parameter must appear with a non-zero coefficient in at least one equation.
Unfortunately, there exist neurons for which these conditions are difficult to satisfy, \textit{i.e.} neurons requiring a very large number of queries to identify a suitable subset of critical points.
Such neurons can be broadly classified into two categories:
\begin{itemize}
   \item[$\bullet$] The first category corresponds to neurons that almost never change state, except on a very small subset of the input space $\mathcal{X}$.
   Finding critical points through random sampling is therefore particularly challenging.
   Examples include \textit{always-off} and \textit{always-on} neurons, as introduced in \autoref{subsec:natural_limit}.
   
   \item[$\bullet$] The second category, which has not been previously reported in the literature, corresponds to neurons for which one of the input entries is almost always null.
   For such neurons, the number of queries required to construct a full-rank matrix can exceed that of standard neurons by a factor of up to $10$ (see \autoref{subsec:extraction_32bits_special_cases}).
   We refer to these neurons as \textit{input-off}.
\end{itemize}

The presence of these neurons is correlated with the type of activation function used.
In particular, the ReLU activation function is known to induce \textit{always-off} neurons~\cite{Dying_ReLU}, and we hypothesize that the repeated application of this activation function is a key factor contributing to the emergence of \textit{input-off} neurons in DNNs.
For these special-case neurons, the signature extraction process must be adapted.
Once a special-case neuron is detected based on the unusually large number of queries required during extraction, a \corr{specific} procedure \corr{must be} applied.
Since these neurons arise in ReLU-based DNNs, we now describe the extraction procedures tailored to each case.

\paragraph{Extraction of \textit{always-off} neurons.} 
The simplest special case is that of an \textit{always-off} neuron.
Since its output is almost always null and therefore independent of the input, its output can be directly set to $0$, without extracting the associated weights.

\begin{figure}[htbp]
    \begin{center}
       \includegraphics[width=0.6\linewidth]{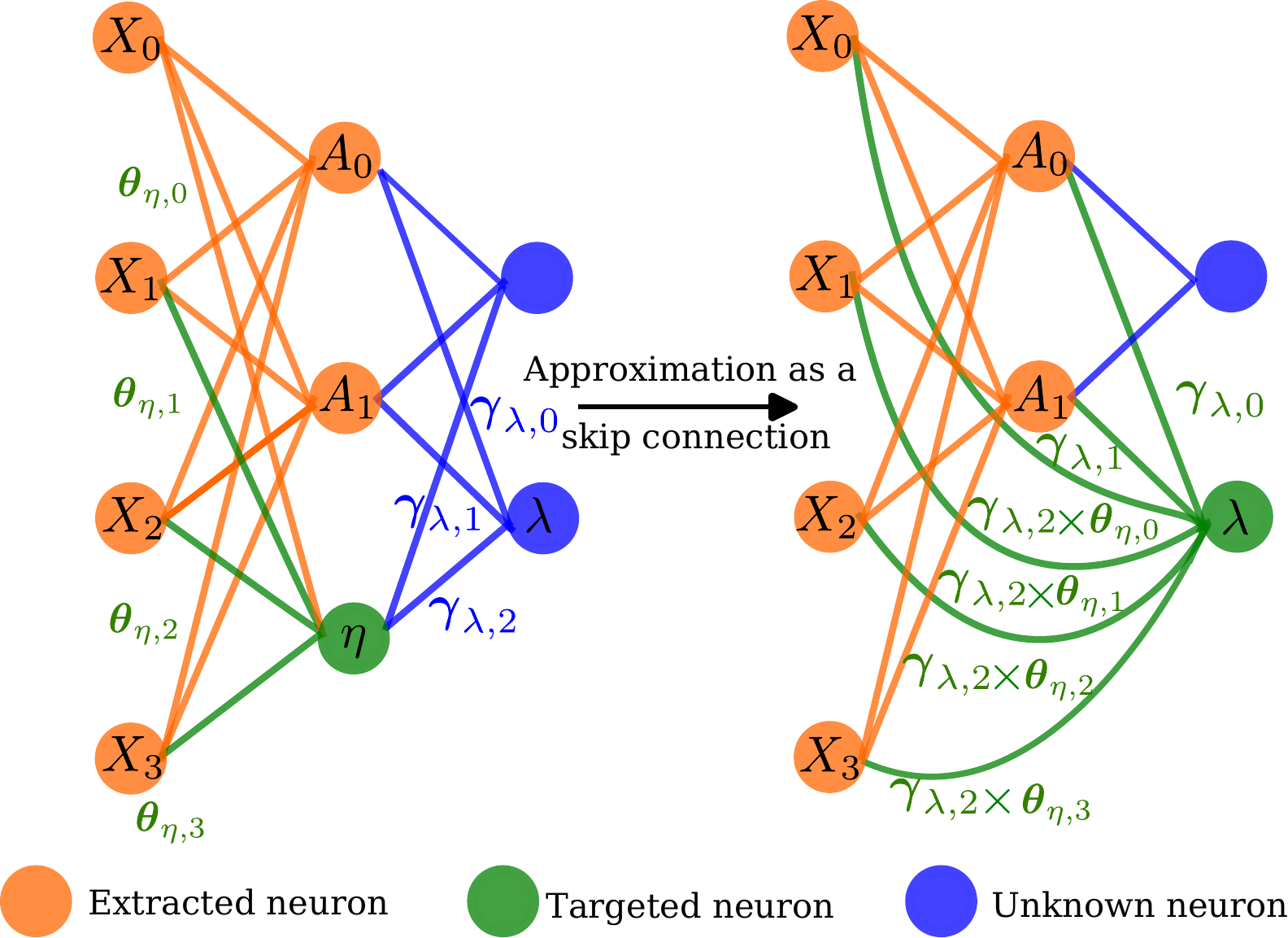}
    \end{center}
    \caption{Approximation of the special-case neuron $\eta$ as a skip connection to the neuron $\lambda$.}
    \label{fig:schema_alwayson}
\end{figure}

\paragraph{Extraction of \textit{always-on} neurons.} 
For \textit{always-on} neurons, the search for critical points can be extremely long and query-intensive.
Moreover, due to the sequential nature of our attack, applying our method to a layer without fully extracting the preceding layers is not possible.
It is therefore essential to extract all neurons at each layer.
Knowing that a neuron is \textit{always-on}, one approach is to treat its activation function, \textit{i.e.} the ReLU, as the identity function and model the neuron as a weighted skip connection between the current layer and the next.
We illustrate this approximation for an \textit{always-on} neuron $\eta$ in the first extracted layer, as part of the signature associated with neuron $\lambda$ in the subsequent layer, in \autoref{fig:schema_alwayson}. 

Let $\boldsymbol{X} \in \mathbb{R}^{n+1}$ be the input to a layer $\ell$ composed of $m$ neurons, and let $\eta$ be an \textit{always-on} neuron in this layer with weight vector $\boldsymbol{\theta}_{\eta} \in \mathbb{R}^{n+1}$ and bias $\beta_{\eta} \in \mathbb{R}$.
Let $\lambda$ denote a neuron in layer $\ell+1$ with weight vector $\boldsymbol{\gamma}_{\lambda} \in \mathbb{R}^{m+1}$ and bias $\beta_{\lambda} \in \mathbb{R}$.
If we denote by $\boldsymbol{A}_{i}$ the activation of neuron $i$ in layer $\ell$, neuron $\eta$ can be approximated as a skip connection to $\lambda$, and \autoref{eq:critical_pts} becomes
\begin{equation}
\label{eq:skip_connection}
\begin{split}
    \sum_{i=0, \atop i\neq \eta}^{m}\boldsymbol{\gamma}_{\lambda, i} \boldsymbol{A}_i + \boldsymbol{\gamma}_{\lambda, \eta} \times \sum_{j=0}^{n}\boldsymbol{\theta}_{\eta, j} \boldsymbol{X}_j + \boldsymbol{\gamma}_{\lambda, \eta}\beta_{\eta} + \beta_{\lambda} &= 0. \\
\end{split}
\end{equation}
For the example shown in \autoref{fig:schema_alwayson}, \autoref{eq:skip_connection} becomes
\begin{equation}
\label{eq:skip_connection_example}
\begin{split}
    \sum_{i=0}^{1}\boldsymbol{\gamma}_{\lambda, i} \boldsymbol{A}_i + \boldsymbol{\gamma}_{\lambda, 2} \times \sum_{j=0}^{3}\boldsymbol{\theta}_{\eta, j} \boldsymbol{X}_j + \boldsymbol{\gamma}_{\lambda, 2}\beta_{\eta} + \beta_{\lambda} &= 0, \boldsymbol{X} \in \mathbb{R}^4. \\
\end{split}
\end{equation} 

As shown above, \autoref{eq:skip_connection} remains linear with respect to the targeted neuron's weights, allowing the construction of a new linear system.
Out of simplicity, we consider that $\eta$ corresponds to the first neuron, \textit{i.e.}, at index $0$, then, using the same notation as before, this system can be expressed as follows:
\begin{equation}
\label{eq:system_skip_connection}
  \left\{
    \begin{aligned}
    -\sum_{i=1}^{m} \frac{\boldsymbol{\gamma}_{\lambda,i}}{\boldsymbol{\gamma}_{\lambda,0}\boldsymbol{\theta}_{\eta,0}}\boldsymbol{A}_i^{1} - \sum_{j=1}^{n} &\frac{-\boldsymbol{\theta}_{\eta,j}}{\boldsymbol{\theta}_{\eta,0}}\boldsymbol{X}_j^{1} - \frac{\beta_{\eta\lambda}}{\boldsymbol{\theta}_{\eta,0}} =\boldsymbol{X}_0^{1}  \\
      &\vdots \\
      -\sum_{i=1}^{m} \frac{\boldsymbol{\gamma}_{\lambda,i}}{\boldsymbol{\gamma}_{\lambda,0}\boldsymbol{\theta}_{\eta,0}}\boldsymbol{A}_i^{z} - \sum_{j=1}^{n} &\frac{-\boldsymbol{\theta}_{\eta,j}}{\boldsymbol{\theta}_{\eta,0}}\boldsymbol{X}_j^{z} - \frac{\beta_{\eta\lambda}}{\boldsymbol{\theta}_{\eta,0}} =\boldsymbol{X}_0^{z}  \\
    \end{aligned}
  \right.
\end{equation}
where $\beta_{\eta\lambda}$ denotes an aggregate of the biases of neurons $\eta$ and $\lambda$, and $z$ is the total number of equations, with $z \geq n + m + 1$.

As this system closely resembles that of \autoref{eq:system_basique}, we solve it using the same approach.
Notably, the signature of neuron $\eta$ appears implicitly in the extracted signature of neuron $\lambda$.
Thus, by solving the LSTSQ problem for a neuron in the subsequent layer, it is possible to recover the signature of an \textit{always-on} neuron.
The only exception is the bias, which cannot be directly separated.
Instead, the solution yields a combination of the biases of $\eta$ and $\lambda$.
Since the weight signatures are already known, only the bias remains unknown.
Consequently, a single additional non-null critical point associated with the targeted \textit{always-on} neuron suffices to extract its bias.

\paragraph{Extraction of \textit{input-off} neurons.}
Finally, the extraction of \textit{input-off} neurons follows a strategy similar to that used for \textit{always-on} neurons.
Here, the signature of a deeper neuron is leveraged to recover the signature of the targeted neuron.
However, unlike \textit{always-on} neurons, \textit{input-off} neurons exhibit state changes depending on the input, preventing the activation function from being approximated as the identity for all inputs.
Since we target ReLU activations, the activation function is either null or the identity.
We therefore rely on the oracle to infer the neuron's state and determine which inputs correspond to each case, allowing us to manually adjust the input equations accordingly.
We then solve the same system of equations as for \textit{always-on} neurons to extract the neuron's signature.
As in the \textit{always-on} case, the bias is recovered using a critical point directly associated with the targeted \textit{input-off} neuron.\\

Using the side-channel oracle, it is therefore possible to extract the signature of a given neuron while increasing robustness to special-case neurons.
However, the signature extraction process assumes that the neuron's state can be reliably inferred and that the specific type of special-case neuron is correctly identified.
\corr{The first assumption is experimentally validated in \autoref{subsec:extraction_32bits_vulnerability}, while the neuron type identification is discussed in \autoref{subsec:extraction_32bits_sign}.
To evaluate the impact of special-case neurons on the model extraction process, an additional analysis is conducted in \autoref{subsec:extraction_32bits_special_cases}, comparing the number of queries required to extract standard and special-case neurons.
Assuming that the signature extraction is correctly performed, a final step remains: estimating the sign of the scaling factor in order to fully extract the $\ell$-th layer.}


\subsection{Estimation of the sign of the scaling factor} \label{subsec:extraction_32bits_sign}
\corr{The extraction of the sign using the \corr{side-channel} oracle $\mathsf{O}$ follows a two-step hypothesis-testing process applied simultaneously to all neurons in the $\ell$-th layer.
First, the neurons in the layer are aligned to reduce the number of hypotheses to a single one for layer $\ell$; second, the correct hypothesis is selected using information from the $(\ell+1)$-th layer.
Furthermore, as discussed in \autoref{subsec:extraction_32bits_signature}, special-case neurons (e.g., \autoref{fig:schema_alwayson}) effectively shift the extraction of a neuron to the $(\ell+1)$-th layer, and therefore require extracting the sign of special-case neurons.
The method proposed in this section is designed to handle these constraints.
}

\paragraph{Layer alignment.} 
The extraction of the sign of all scaling factors begins once all neuron signatures in the \corr{$\ell$-th layer} have been extracted. 
At this stage, the attacker knows the induced hyperplane for each neuron and can arbitrarily assign a state to each side of the hyperplane, denoted $V_A$ and $V_B$, resulting in $2^n$ possible hypotheses, where $n$ is the number of neurons in the layer.
To reduce this number, the neurons are aligned such that, for each neuron in the \corr{layer $\ell$}, the arbitrary state $V_A$ corresponds to the same real neuron state (either active or inactive), while $V_B$ corresponds to the opposite state.
This reduces the number of hypotheses to a single candidate corresponding to the identification of the sign associated with $V_A$ or $V_B$.
This alignment is performed using the \corr{side-channel oracle} $\mathsf{O}$.
Indeed, by comparing the states of two neurons $\eta$ and $\gamma$ for a given input $\boldsymbol{X}$ (\textit{i.e.} $V(\eta; \boldsymbol{X})$ and $V(\gamma; \boldsymbol{X})$), it is possible to align the scaling factors associated with the signatures of these two neurons.

\paragraph{Hypothesis selection.} 
\corr{Once the neurons of the $\ell$-th layer have been aligned, an arbitrary assumption is made regarding the states $V_A$ and $V_B$; for instance, $V_A$ (resp. $V_B$) is assumed to correspond to the active (resp. inactive) state.
The validity of this hypothesis is then evaluated during the signature extraction of the first neuron in the $(\ell+1)$-th layer.
This approach is justified as follows: once a sufficient number of critical points has been extracted in the $(\ell+1)$-th layer, a linear relationship is expected between these points and the output of the $\ell$-th layer (see \autoref{subsec:extraction_32bits_signature}).
An incorrect assumption on $V_A$ and $V_B$ breaks this linear relationship, as the ReLU function computes $\boldsymbol{A} = \texttt{ReLU}(-\boldsymbol{X})$ instead of $\boldsymbol{A} = \texttt{ReLU}(\boldsymbol{X})$.
This results in incorrect activation values, \textit{i.e.} incorrect inputs to the $(\ell+1)$-th layer, such that \autoref{eq:critical_pts} no longer holds.
Consequently, the residual error of the LSTSQ solution (see \autoref{subsec:extraction_32bits_signature}) is used as a criterion to validate the sign hypothesis.
The correct hypothesis is the one that yields the smallest residual error.} \\

Using this method, we can extract the sign of each neuron without requiring additional queries.
Furthermore, since the method requires only one hypothesis per layer, it remains practical even for complex and deep architectures.
\corr{The final step that must be justified is the ability to retrieve the sign of special-case neurons.}

\paragraph{Identifying the type of special-case neuron.} 
\corr{By definition, special-case neurons cannot appear in the first layer of the targeted DNN, since the attacker controls the inputs sent to the network.
Consequently, when a special-case neuron $\eta$ is detected in layer $\ell$, the preceding layer $(\ell-1)$ has already been fully extracted.
Because this previous layer is known, the sign of its scaling factors is also known.
The attacker can therefore select any neuron $\gamma$ in layer $(\ell-1)$ and use the side-channel oracle $\mathsf{O}$ to compare the states of neurons $\eta$ and $\gamma$ for a given input.
Depending on the oracle’s output, the attacker can directly determine the state of the special-case neuron and, as a result, identify its type.} \\

With this oracle, it is therefore possible to extract the critical points, the signature, and the sign of the scaling factor while allowing the extraction of special-base neurons.
\corr{In the next section, we demonstrate how this divide-and-conquer methodology can be used to extract pooling layers for the first time.}


\subsection{\corr{Discussion on the methodology}}
\label{subsec:discuss_methodo}

\subsubsection{Pooling layer extraction}
\begin{figure}[htbp]
    \begin{center}
       \includegraphics[width=0.7\linewidth]{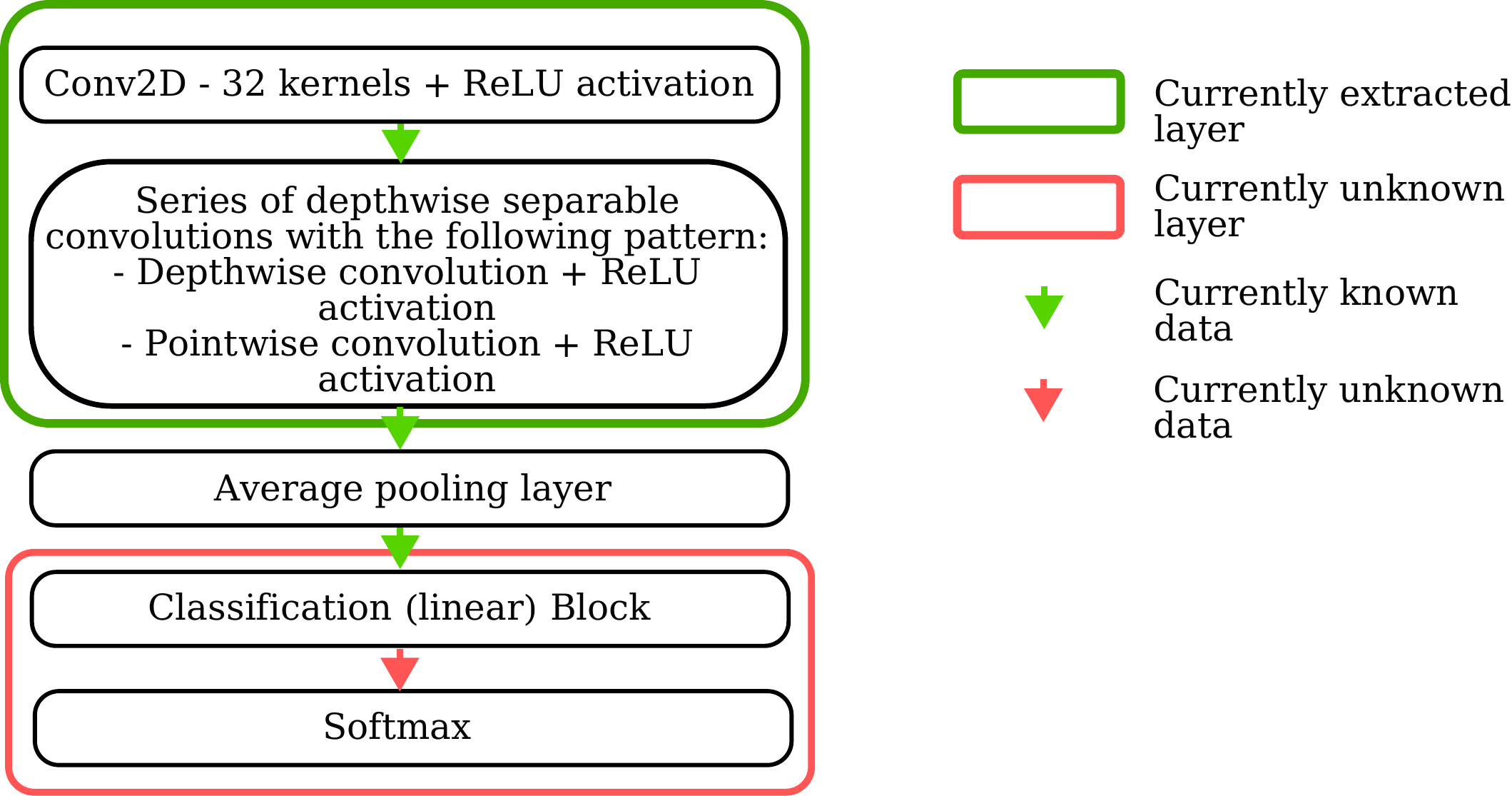}
    \end{center}
    \caption{Extraction process of an architecture including an average pooling layer. (\textbf{Note}: It is assumed that all depthwise separable convolutions have been successfully extracted using our methodology).}
    \label{fig:split_MobileNetv1}
\end{figure}

Under the assumption that an attacker can extract the signatures and the signs of the scaling factors using the divide-and-conquer methodology, it becomes possible to target modern architectures composed of non–fully connected layers, such as MobileNets~\cite{Howard2017MobileNetsEC}.
In \autoref{fig:split_MobileNetv1}, we provide a high-level description of the architecture of \textit{MobileNetv1}.
This model consists of a sequence of depthwise separable convolutions, followed by an average pooling layer, and finally a classification block containing a fully connected layer and a softmax function.
Due to the presence of the average pooling layer, such architectures have not previously been targeted by cryptanalytic-based methods~\cite{Jagielski2019HighAA, Rolnick2019ReverseengineeringDR, Carlini2020CryptanalyticEO}, as discussed in \autoref{subsec:natural_limit}.\\

\corr{Based on the methodology introduced in \autoref{subsec:divide_conquer_methodo}, an attacker can leverage the side-channel oracle to target convolutional operations.
As with fully connected layers, this process consists of targeting the state of each neuron induced in the depthwise and pointwise convolutional layers.
Since the convolutional layers can be represented as a special kind of fully connected layer with a very particular sparsity and weight-sharing pattern, the divide-and-conquer methodology can be straightforwardly applied in this setting.
Additionally, as activation functions are used in such layers, the neuron's state can be retrieved equivalently.}
Moreover, in MobileNetv1, the pooling layer is placed immediately after an activation function, which allows its impact to be ignored when using the oracle.
To illustrate this point, consider the case where all preceding layers have been successfully extracted.
As shown in \autoref{fig:split_MobileNetv1}, the input to the average pooling layer is then known; because this layer is non-parametric, its output is also known, and the attacker can directly target the subsequent layer.
Consequently, pooling layers can be disregarded in such configurations, as they correspond to known transformations applied to known data.
This demonstrates that complex architectures, including those with pooling layers such as MobileNetv1, can be effectively targeted.
\corr{While we have described the benefits of using the side-channel oracle to target more complex architectures, the following section discusses the choice of oracle and argues that the divide-and-conquer strategy could benefit from other types of oracles.}


\subsubsection{\corr{Choice of the oracle}}
\corr{Our methodology relies on an oracle $\mathsf{O}$ that returns a boolean value indicating whether two neurons are associated with the same state for two given inputs.
Any function achieving this goal can be used as an oracle within our framework.
As a result, the methodology is not restricted to side-channel attacks or to the ReLU activation function.
For instance, one could design an oracle based on fault-injection techniques by testing the activity of a neuron through the absence or presence of variations in the DNN output after faulting the pre-activation value, in a manner similar to the approach proposed in~\cite{Hector2023FaultIA} for weight extraction.
Similarly, if an attacker targeting other activation functions, such as tanh or ReLU6, is able to construct a function that distinguishes between different activation states, then our methodology remains applicable.
It is worth noting that the activation states do not necessarily need to be binary or localized at $0$, as is the case for the ReLU function.
For example, in the case of ReLU6, if the attacker can distinguish between the inactive state (\textit{i.e.} negative pre-activation values), the identity state (\textit{i.e.} activation values between $0$ and $6$), and the constant state (\textit{i.e.} activation values greater than $6$), the methodology can be adapted by incorporating additional equations derived from points located between the identity and constant states.
In such a scenario, the threat model would no longer be restricted to DNNs using the ReLU activation function.
We leave the exploration of alternative oracles and activation functions for future work.}
\section{Practical experiments}\label{sec:extraction_32bits_experiments}

In this section, we test our methodology by first validating our side-channel-based oracle and then performing the extraction of DNNs.
We begin with a description of the targeted architectures in \autoref{subsec:extraction_32bits_setup}.
Then, in \autoref{subsec:extraction_32bits_vulnerability}, we describe the vulnerability in the ReLU function and evaluate the effectiveness of our side-channel-based oracle.
Our results on classifiers are presented in \autoref{subsec:extraction_32bits_result_classifiers}, and we provide a comparison between our methodology and soft-label extraction methods in \autoref{subsec:extraction_32bits_result_simulation}.
Finally, in \autoref{sec:extraction_32bits_discussion}, we discuss the perspectives and limitations of our results.

\subsection{Experimental setup}\label{subsec:extraction_32bits_setup}

\paragraph{Dataset and Architecture.}
We target several embedded DNNs, including a shortened MobileNetv1 \cite{Howard2017MobileNetsEC}.
A detailed description of this model is provided in \autoref{tab:extraction32bits_archi_target_mobilenetv1}.
This DNN shares the same basic blocks as the original MobileNetv1, namely depthwise separable convolutions.
The architecture of a depthwise separable convolution is described in \autoref{tab:archi_DW_sep}.
To the best of our knowledge, these blocks have never been successfully extracted using cryptanalytic extraction methods.
We also tested our methodology on a large MLP with the architecture described in \autoref{tab:archi_bigMLP}.

For clarity, when describing MLPs, we adopt the following notation: Size Input - (Size Hidden Layer) - Size Output Layer.
For instance, the large MLP described above is denoted as $3072$-$512$-$256$-$64$-$10$.
Finally, we target the most complex architectures considered in \cite{Jagielski2019HighAA, Rolnick2019ReverseengineeringDR, Carlini2020CryptanalyticEO, Chen2024HardLabelCE, Carlini2024PolynomialTC}.
A summary of the targeted architectures is provided in \autoref{tab:all_targets}.

\begin{table}[ht]
\caption{Summary of the targeted architectures.}
\label{tab:all_targets}
\centering
\resizebox{\linewidth}{!}{
\begin{tabular}{|p{0.35\linewidth}|p{0.25\linewidth} | p{0.15\linewidth} | p{0.25\linewidth}|}
\hline
\textbf{Type of architecture} & 
\textbf{Nb. of parameters} & 
\textbf{Nb. of targets} & 
\textbf{Targeted by} \\
\hline
\textbf{Small MLP} &  $< 200,000$ & $7$ & \cite{Carlini2020CryptanalyticEO, Jagielski2019HighAA, Rolnick2019ReverseengineeringDR, Chen2024HardLabelCE}, This work \\
\hline
\textbf{MLP} & $935,370$ & $1$ & \cite{Carlini2024PolynomialTC}, This work \\
\hline
\textbf{Large MLP} & $1,721,802$ & $1$ & This work \\
\hline
\textbf{MobileNetv1-short} & $5,234$ & $1$ & This work \\
\hline
\end{tabular}}
\end{table}

All classifiers are trained on the CIFAR-10 dataset\footnote{\url{https://www.cs.toronto.edu/~kriz/cifar.html}} except for the architecture from \cite{Chen2024HardLabelCE}.
This network, as well as all DNNs performing regression, is trained on a random dataset, as in the SOTA.

\paragraph{Hardware configuration.} 
The models are deployed on an STM32-F767ZI board, based on an ARM Cortex-M7 MCU, via the X-Cube-AI framework without compiler optimizations, using $32$-bit floating-point data for all values, \textit{e.g.} inputs, weights, and activation values.
The board features $2$ MBytes of Flash memory and $512$ KBytes of SRAM.
X-Cube-AI is a framework developed by ST Microelectronics to facilitate DNN deployment on STM's embedded devices.
However, this framework is not designed to provide secure deployment of embedded DNNs.

\paragraph{Practical setup.} 
An EM probe from Langer (EMV-Technik RF-U 2.5) is connected to an amplifier (ZLF-2000G+) to capture EM traces.
A Lecroy oscilloscope (2.5 GHz WaveRunner 625Zi) is used to acquire the EM signal.
To simplify leakage exploitation, a trigger surrounding the ReLU processing is added.
Since the patterns around the ReLU function are distinguishable from those of the preceding matrix product, the trigger configuration is not considered difficult and does not alter the threat model considered in this paper.

\subsection{Side-channel analysis of the ReLU implementation} \label{subsec:extraction_32bits_vulnerability}

Our methodology relies entirely on the oracle's ability to distinguish the state of two inputs for a given neuron.
Thus, we first validate that such a function can be constructed using the distinguisher introduced in \autoref{subsec:divide_conquer_methodo}.
Here, we present results obtained using the setup described above.

\paragraph{Description of the vulnerability.} 
The ReLU implementation in X-Cube-AI is constant-time, as recommended in \cite{Maji2021LeakyNR}.
\autoref{alg:ReLU_implem} illustrates the main idea for $32$-bit values\footnote{Understanding the vulnerability is not required to perform our black-box extraction attack; we present it here for illustration purposes.}.
\begin{algorithm}
\caption{Pseudo-code of the constant-time ReLU function.}\label{alg:ReLU_implem}
\begin{algorithmic}
\Require Pre-activation value $z \in \mathbb{R}$
\Ensure $a = \texttt{max}(0,z) \in \mathbb{R}^{+}$
\State $a \gets (\sim (z \gg 31))$ \& $ z$ 
\end{algorithmic}
\end{algorithm}
For a signed pre-activation value $z \in \mathbb{R}$ encoded on $32$ bits, the sign can be extracted via $(z\gg 31)$ such that it is then cast to 32 bits to serve as a mask, \textit{i.e.} $(\sim (z \gg 31)) \in \{\texttt{0x00000000, 0xFFFFFFFF}\}$.
Depending on the sign of $z$, applying this mask either leaves the value unchanged or sets the output to zero.
Although the implementation is constant-time, the mask operation produces a physical leakage exploitable by an attacker.
Specifically, the mask is either $\texttt{0x00000000}$ or $\texttt{0xFFFFFFFF}$ depending on the neuron's state.
The difference in Hamming weights between these two values allows separation into two groups using clustering algorithms, corresponding to the active and inactive states.

\paragraph{Critical point search.} 
To validate our method, we acquire $10,000$ EM traces using the setup in \autoref{subsec:extraction_32bits_setup}.
We then follow the strategy in \autoref{subsec:divide_conquer_methodo} and measure the success rate of retrieving the state of $12$ neurons.
The \textit{k-means} algorithm is used to cluster the neuron states without prior knowledge of their true states.
As noted in \autoref{subsec:divide_conquer_methodo}, only correct separation between the two clusters is required for the attack.
We compute the success rate by comparing the cluster assignments to the true neuron states, \textit{i.e.} verifying that neurons with the same state are assigned to the same cluster.
Using one EM trace, we achieve an average success rate of $85.9\%$.
While this validates our methodology, high query complexity (as in \cite{Jagielski2019HighAA}, see \autoref{subsec:extraction_32bits_split}) motivates achieving near-100\% success for complex DNNs.
A simple way to improve success is to acquire multiple traces for the same input and use majority voting to infer the state.

\begin{figure}[ht]
\centering
    \begin{subfigure}{.48\linewidth}
        \includegraphics[width=\linewidth]{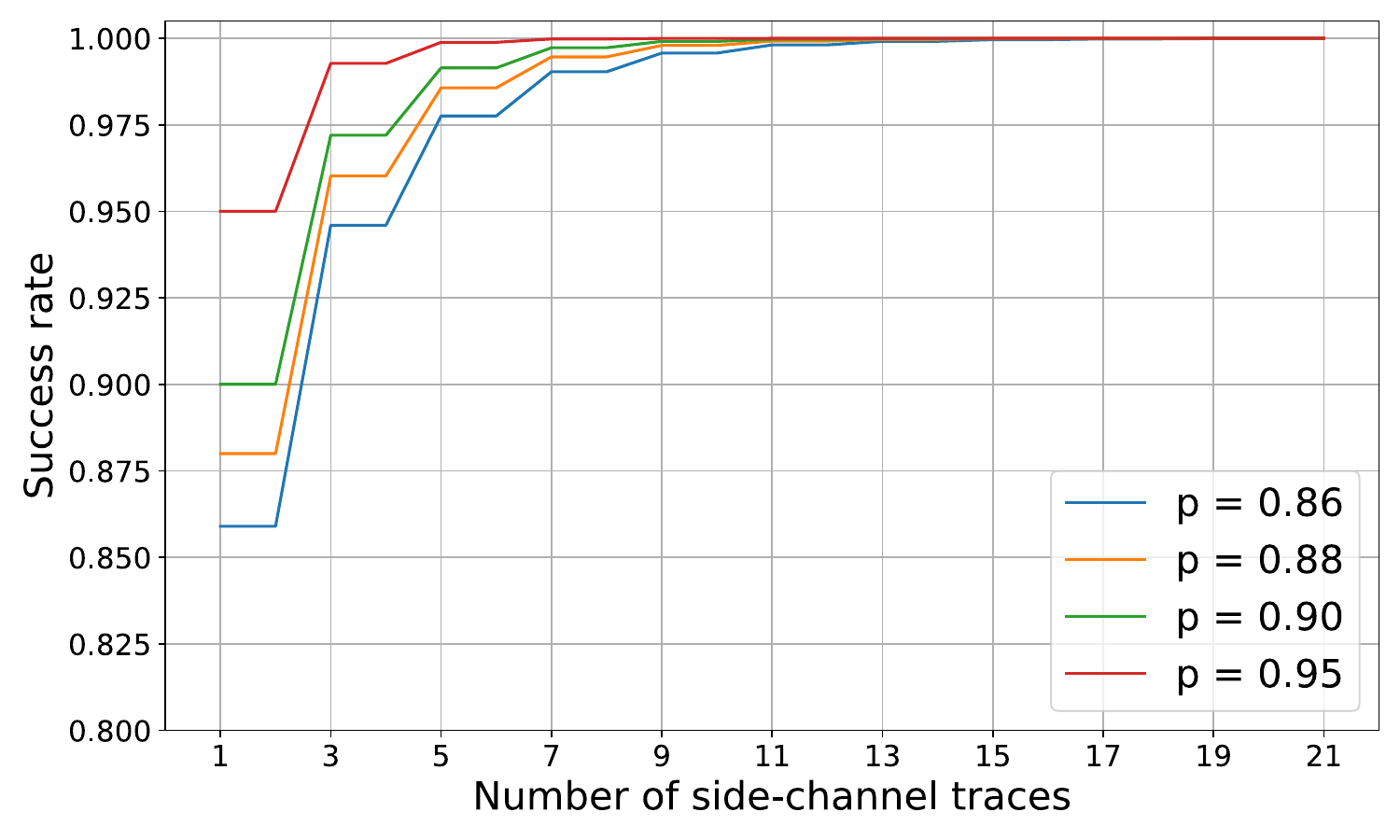}
        \caption{Success rate with different initial probabilities.}
        \label{fig:success_rate_vs_ntraces}
    \end{subfigure}%
    \hfill
    \begin{subfigure}{.48\linewidth}
        \includegraphics[width=\linewidth]{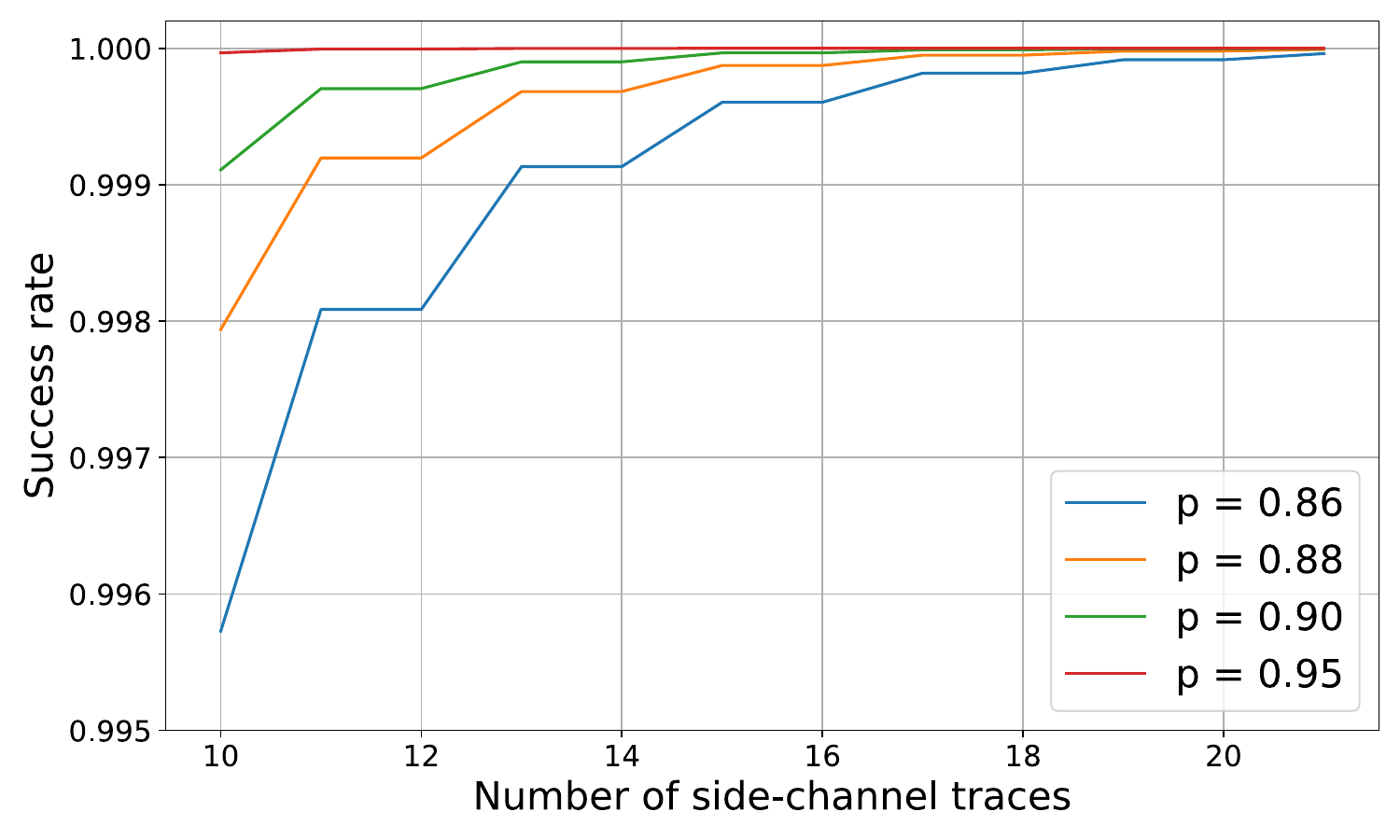}
        \caption{Zoom of \autoref{fig:success_rate_vs_ntraces} after ten traces in the same settings.}
        \label{fig:success_rate_zoom}
    \end{subfigure}
    \caption{Evolution of the success rate of neuron state extraction as a function of the number of side-channel traces.}
    \label{fig:evolution_success_rate}
\end{figure}

\autoref{fig:success_rate_vs_ntraces} shows the success rate obtained using majority voting versus the number of traces for different initial success probabilities.
With an initial success of $85.9\%$, the success rate exceeds $99\%$ with $7$ traces and $99.9\%$ with $13$ traces.
For worst-case analysis, we assume an ideal scenario where one trace suffices for a 100\% success rate.
All results in the following sections are reported under this assumption, providing an empirical upper bound for performance.
This assumption is not restrictive, since if $N$ traces are required to achieve 100\% success, the overall complexity scales linearly with $N$.
\corr{Finally, as these results depend on the targeted device, \textit{i.e.} other devices may produce lower success rates, we discuss the impact of oracle error rates in \autoref{subsec:extraction_32bits_impact_error_rate}.}

\paragraph{Sign extraction.} 
We also validate that the oracle can align neuron states for sign extraction.
Using the method in \autoref{subsec:extraction_32bits_sign}, the layer-level clusters are reduced to two.
This achieves a success rate of $99.7\%$ for retrieving the sign of each neuron's scaling factor from a single trace.
Our results are independent of the clustering algorithm and could be improved using other unsupervised methods.
Since the vulnerability originates from the software implementation, any platform using it is a potential target.
To our knowledge, this represents the first successful extraction of neuron scaling factor signs in a hard-label setting without additional queries.

\subsection{Complete extraction of classifiers} \label{subsec:extraction_32bits_result_classifiers}

\paragraph{Evaluation metrics.} 
To evaluate the fidelity of the extracted models, we compute the metrics introduced in \cite{Rakin2021DeepStealAM} and \cite{Hector2023FaultIA}: \textit{Fidelity} and \textit{Accuracy Under Attack} (AuA).
Fidelity metric measures the average label agreement between the \corr{copied} model and the original model on the \textit{top-1} prediction for a subset of inputs.
AuA measures the transfer rate of adversarial examples generated on the \corr{copied} model and tested on the targeted model, serving as a proxy for the ability to craft effective attacks on the victim's network using the \corr{copied} model.
We also compute the maximum absolute difference between the true weights $\theta$ and their extracted counterparts $\hat{\theta}$ to characterize extraction success, as in \cite{Jagielski2019HighAA, Rolnick2019ReverseengineeringDR, Carlini2020CryptanalyticEO}.
This metric is computed both on the entire network of depth $L$ ($\texttt{max}|\Delta_{\theta}|^{L}$) and on the first $L-1$ layers ($\texttt{max}|\Delta_{\theta}|^{L-1}$).
Finally, the number of queries required to the targeted model is used to evaluate the cost of the attack.

\paragraph{Summary of the complete results.} 
To our knowledge, cryptanalytic extraction of DNNs in hard-label settings has previously only been performed in \cite{Chen2024HardLabelCE, Carlini2024PolynomialTC}. 
We first test our methodology on the largest architectures from these works: a 1024-2-2-1 MLP \cite{Chen2024HardLabelCE} and a 3072-256-256-256-64-10 MLP \cite{Carlini2024PolynomialTC}, referred to as MLP-a and MLP-b, respectively.
It is important to note that while \cite{Carlini2024PolynomialTC} extracts the hidden layers, it does not extract the last layer, making it impossible to report metrics for the full network, \textit{i.e.}, number of queries, fidelity, and $\texttt{max}|\Delta_{\theta}|^{L}$.
We additionally extract a 3072-512-256-64-10 MLP (MLP-c) with $1.7$ million parameters, nearly doubling the previous maximum number of parameters extracted in SOTA studies.
The results are summarized in \autoref{tab:sota_classifier}.

\begin{table*}[htbp]
\caption{Comparison in hard-label settings.}
\label{tab:sota_classifier}
\centering
\resizebox{\linewidth}{!}{
\begin{tabular}{|p{0.35\linewidth}| p{0.13\linewidth}| p{0.15\linewidth}| p{0.1\linewidth} | p{0.12\linewidth}| p{0.12\linewidth} | p{0.12\linewidth}|}
\hline
\textbf{Architecture} & \textbf{Parameters} & \textbf{Approach} & \textbf{Queries} & \textbf{$\texttt{max}|\Delta_{\theta}|^{L-1}$}& \textbf{$\texttt{max}|\Delta_{\theta}|^{L}$} & \textbf{Fidelity}\\
\hline
\textbf{MLP-a 1024-2-2-1} & \multirow{2}{*}{\textbf{2,059}} & \cite{Chen2024HardLabelCE} & $ 2^{22.49}$ & - &$ \boldsymbol{2^{-20.38}}$ & $\boldsymbol{100\%}$\\
\textbf{(64-bit data)} &  & \textbf{This work} & $ \boldsymbol{2^{17.12}}$ & $\boldsymbol{2^{-42.5}}$ & $ 2^{3.45}$ & $99.5\%$\\
\hline
\textbf{MLP-b 3072-256-256-256} & \multirow{2}{*}{\textbf{935,370}} & \cite{Carlini2024PolynomialTC} & - &  $ 2^{-18.0}$ & - & - \\
\textbf{-64-10 (64-bit data)} &  & \textbf{This work} & $ \boldsymbol{2^{26.2}}$ & $\boldsymbol{2^{-26.6}}$ & $ 2^{2.1}$ & $97.2\%$\\
\hline
\textbf{MLP-c 3072-512-256-64-10 (32-bit data)} & \textbf{1,721,802} & \textbf{This work} & $2^{26.0}$ & $2^{2.3}$ & $2^{3.2}$ & $93.2\%$\\
\hline
\textbf{Truncated MobileNetv1 (32-bit data)} & \textbf{5,234} & \textbf{This work} & $2^{18.6}$ & $2^{-4.3}$ & $2^{3.7}$ & $88.4\%$\\
 \hline
\end{tabular}}
\end{table*}

\paragraph{Performance comparison.} 
We successfully extract the architectures mentioned above, achieving $99.5\%$ fidelity for MLP-a and $97.2\%$ for MLP-b. 
Compared to \cite{Chen2024HardLabelCE}, our methodology reduces the number of queries by a factor of $\approx 41$ while maintaining comparable fidelity ($100\%$ in \cite{Chen2024HardLabelCE} vs $99.5\%$ in our work).
This drop in the fidelity is due to the fact that their methodology is specifically designed for binary classifiers.
Indeed, for binary classifiers, it is possible to form a system of equations using decision boundary points, \textit{i.e.}, points on the boundary where the output label changes, and the activation value of the previous layer. 
While this allows for precise extraction of the last layer, it is not directly applicable to all multiclass classifiers.
Therefore, we use our method, which is more generic.

Since Carlini \textit{et al.} \cite{Carlini2024PolynomialTC} do not perform complete extraction of MLP-b, we can only evaluate the maximum error on the hidden layer weights.
Using this metric, our method achieves a more accurate estimation by a factor of $388$.
In addition, \cite{Carlini2024PolynomialTC} assumes perfect extraction of the previous layer when targeting a new one, removing residual error propagation.
For comparison, we simulated our methodology under the same assumption and computed the maximum absolute difference with and without residual error, as shown in \autoref{tab:extraction32bits_no_error}.
\begin{table}[ht]
\caption{Maximum error on the weights with and without the residual error on the MLP architecture from \cite{Carlini2024PolynomialTC}}
\label{tab:extraction32bits_no_error}
\centering
\resizebox{\linewidth}{!}{
\begin{tabular}{|p{0.35\linewidth}|p{0.13\linewidth} | p{0.13\linewidth} | p{0.13\linewidth} | p{0.13\linewidth}| p{0.13\linewidth}|}
\hline
\textbf{Assumption} & \textbf{Layer 1} &  \textbf{Layer 2} & \textbf{Layer 3} & \textbf{Layer 4} & \textbf{Layer 5}\\
\hline
With residual error & $2^{-38.1}$ & $2^{-33.8}$ & $2^{-29.7}$ & $2^{-26.6}$ & $2^{2.1}$ \\
\hline
Without residual error & $2^{-38.1}$ & $2^{-41.9}$ & $2^{-42.8}$ & $2^{-41.3}$ & $2^{2.0}$ \\
\hline
\end{tabular}}
\end{table}

Under the assumption of no error propagation, performance is consistent across layers, and our method further improves the results of \cite{Carlini2024PolynomialTC}, reducing the maximum hidden layer weight error by a factor of $\approx 2^{20}$.
Finally, even when considering error propagation, we achieve high-fidelity extraction of the new architectures: $88.4\%$ for MobileNetv1 and $93.2\%$ for MLP-c on the full CIFAR-10 test set ($10,000$ images).

\paragraph{Comparison with \textit{learning-based} extraction.} 
We compare our results with \textit{learning-based} extraction methods using a baseline model trained on a balanced dataset composed of the queries used during the critical points' search, with the responses of the targeted model as labels.
The baseline model uses the same hyperparameters as the targeted model, \textit{e.g.}, architecture and learning rate.
While the baseline approach achieves reasonable performance during training (around $56\%$ accuracy), it generalizes poorly, and its performance drops significantly on the test set, achieving only $21.1\%$ fidelity for MobileNetv1.
The use of random queries, which do not follow the same distribution as the true CIFAR-10 training data, combined with the hard-label setting, drastically reduces the baseline model's performance.
This illustrates that \textit{learning-based} extraction methods are heavily dependent on data distribution, highlighting the advantage of cryptanalytic attacks for achieving higher fidelity in the threat model considered in this work.

\paragraph{Generation of adversarial examples.} 
To complete our performance evaluation, we use AuA to quantify how much our extracted model can improve other attacks, particularly adversarial attacks.
We generate adversarial examples using the \corr{copied} model and measure their impact on the targeted model, employing three different attack algorithms.
\corr{We first use the PGD attack \cite{madry2018towards} with the settings from \cite{Rakin2021DeepStealAM, Hector2023FaultIA}, \textit{i.e.} a perturbation factor of $8/255$ and $40$ iterations.
We also generate adversarial examples using the BIM attack \cite{kurakin2018adversarial} with up to $4$ iterations and a gradient step of $5\times10^{-3}$.
Finally, we use the C$\&$W attack \cite{carlini2017towards} with the L2 norm, using the CleverHans implementation \cite{papernot2018cleverhans}, a learning rate of $5\times10^{-3}$, and a maximum of $1000$ iterations.
All the results are reported in \autoref{tab:adv_examples}.
For MLP-b, MLP-c and MobileNetv1, the transfer rates are close to white-box performance across all evaluated methods.}
These results outperform a powerful substitute model trained with access to the true data \cite{Rakin2021DeepStealAM}, demonstrating that our methodology enables more effective attacks in hard-label settings.

\begin{table}[ht]
\caption{Transfer rates (in term of AuA) of adversarial examples generated on the copied model and tested on the targeted model.}
\label{tab:adv_examples}
\centering
\resizebox{\linewidth}{!}{
\begin{tabular}{|p{0.55\linewidth}|p{0.13\linewidth} | p{0.13\linewidth} | p{0.13\linewidth} |}
\hline
\textbf{Architecture} & \textbf{PGD} \cite{madry2018towards} &  \textbf{BIM} \cite{kurakin2018adversarial} & \textbf{C\&W} \cite{carlini2017towards}\\
\hline
\textbf{MLP-b 3072-256-256-256-64-10 (64-bit data)} & $98.04 \%$ & $98.71 \%$ & $89.61 \%$ \\
\hline 
\textbf{MLP-c 3072-512-256-64-10 (32-bit data)} & $97.48 \%$ & $98.40 \%$ & $90.27 \%$ \\
\hline
\textbf{Truncated MobileNetv1 (32-bit data)} & $98.62 \%$ & $97.30 \%$ & $98.01 \%$ \\
\hline
\end{tabular}}
\end{table}

\subsection{Simulation of our methodology in soft-label setting} \label{subsec:extraction_32bits_result_simulation}

To ensure a fair comparison between our methodology and other cryptanalysis-based approaches, we include results from SOTA on DNNs trained for regression tasks, assuming access to the continuous output value.
In this context, we leverage this access to improve the extraction of the last layer, which differs from the previous layers, as mentioned in \autoref{subsec:divide_conquer_methodo}.
In contrast, previous methodologies use this output to extract the hidden layers directly and can perform gradient-based optimizations to improve extraction efficiency.
However, none of these works target DNNs in hard-label settings, unlike our methodology.

\paragraph{Evaluation metrics.} 
For regression tasks, metrics such as AuA and fidelity, \textit{i.e.}, label agreement, are no longer meaningful. 
Consequently, SOTA methodologies evaluate extraction fidelity by considering only the maximum error on the extracted weights \cite{Jagielski2019HighAA, Rolnick2019ReverseengineeringDR,Carlini2020CryptanalyticEO}.
This metric has the advantage of providing an estimate of fidelity that is independent of any assumptions about the input distribution, unlike a layer-wise mean squared error.
Finally, we continue to use the number of queries to quantify the cost of the attack. \\

As our methodology is not based on the gradient of the targeted DNN, it is not affected by the precision of the output format.
Previous methodologies target DNNs using $64$-bit data.
In this section, we first present results obtained from simulations using $64$ bits and then provide results using $32$-bit data.
We use the same notation for describing the architecture as introduced in \autoref{subsec:extraction_32bits_setup}, \textit{i.e.}, a 784-128-1 DNN corresponds to a model with an input size of $784$, one hidden layer composed of $128$ neurons, and one output.

\paragraph{Results on DNN using $64$-bit data.} 
\autoref{tab:res_sota_64} provides a comparison of DNNs trained for regression in soft-label settings with 64-bit data. In terms of extraction fidelity, our methodology outperforms all previous methods, reducing the maximum error on the weights by factors ranging from $2^{4.5}$ to $2^{16.1}$, despite operating in hard-label settings for the hidden layers. 
The side-channel attack allows for more precise localization of critical points compared to gradient-based methods, providing a more accurate estimation of the weights.
Regarding attack cost, our methodology requires fewer queries for four out of the six previously targeted architectures. 
For the two architectures where our attack is more costly, we hypothesize that this is due to their narrow topology, \textit{i.e.} only one hidden layer, which benefits gradient-based optimizations used in SOTA methodologies.
For the remaining architectures, the direct mapping from critical points to neurons provided by the side-channel attack improves our methodology's efficiency relative to SOTA.

\begin{table}[ht]
\caption{Comparison of DNNs trained for regression in soft-label settings with 64-bit data.}
\label{tab:res_sota_64}
\centering
\resizebox{\linewidth}{!}{
\begin{tabular}{|p{0.25\linewidth}|p{0.2\linewidth}| p{0.2\linewidth} | p{0.12\linewidth} | p{0.19\linewidth}|}
\hline
\textbf{Architecture} & \textbf{Parameters} & \textbf{Approach} & \textbf{Queries} & \textbf{max}$|\boldsymbol{\Delta_{\theta}|^{L}}$ \\
\hline
\multirow{3}{*}{\textbf{10-10-10-1}} & \multirow{3}{*}{$210$} & \cite{Carlini2020CryptanalyticEO} &$2^{16.0}$ & $2^{-36.0}$\\
 & & \cite{Rolnick2019ReverseengineeringDR} &$2^{22.0}$ & $2^{-12.0}$\\
 & & \textbf{This work} &$ \boldsymbol{2^{15.6}}$ & $\boldsymbol{2^{-46.2}}$ \\
\hline
\multirow{2}{*}{\textbf{10-20-20-1}} & \multirow{2}{*}{$620$} & \cite{Carlini2020CryptanalyticEO} &$2^{17.1}$ & $2^{-37.0}$\\
 & & \textbf{This work} &$ \boldsymbol{2^{15.6}}$ & $\boldsymbol{2^{-46.5}}$ \\
\hline
\multirow{2}{*}{\textbf{40-20-10-10-1}} & \multirow{2}{*}{$1,110$} & \cite{Carlini2020CryptanalyticEO} &$2^{17.8}$ & $2^{-27.1}$\\
 & & \textbf{This work} &$\boldsymbol{2^{16.8}}$ & $\boldsymbol{2^{-42.0}}$ \\
\hline
\multirow{2}{*}{\textbf{80-40-20-1}} & \multirow{2}{*}{$4,020$} & \cite{Carlini2020CryptanalyticEO} &$2^{18.5}$ & $2^{-39.7}$\\
 & & \textbf{This work} &$ \boldsymbol{2^{18.3}}$ & $\boldsymbol{2^{-44.2}}$ \\
\hline
\multirow{3}{*}{\textbf{784-32-1}} & \multirow{3}{*}{$25,120$} & \cite{Carlini2020CryptanalyticEO} &$2^{19.2}$ & $2^{-30.2}$\\
 & & \cite{Jagielski2019HighAA} &$\boldsymbol{2^{18.2}}$ & $2^{-1.7}$\\
 & & \textbf{This work} &$ 2^{20.6}$ & $\boldsymbol{2^{-43.5}}$ \\
\hline
\multirow{2}{*}{\textbf{784-128-1}} & \multirow{2}{*}{$100,480$} & \cite{Carlini2020CryptanalyticEO} &$\boldsymbol{2^{21.5}}$ & $2^{-24.7}$\\
 & & \textbf{This work} &$ 2^{22.6}$ & $\boldsymbol{2^{-40.8}}$ \\
\hline
\end{tabular}}
\end{table}

\paragraph{Results on DNN using $32$-bit data.} \autoref{tab:res_sota_supp_32} provides a comparison of our methodology against targeted architectures from SOTA methodologies using $32$-bit data. 
Although no prior comparisons exist in this setting, we present these results for future reference.
We successfully extract all architectures under these conditions as well.

\begin{table}[ht]
\caption{Results of our methodology against targeted architectures by SOTA methodologies using $32$-bit data.}
\label{tab:res_sota_supp_32}
\centering
\resizebox{\linewidth}{!}{
\begin{tabular}{|p{0.3\linewidth}|p{0.25\linewidth}| p{0.2\linewidth} | p{0.25\linewidth}|}
\hline
\textbf{Architecture} & \textbf{Parameters}  & \textbf{Queries} & \textbf{max$|\Delta_{\theta}|^{L}$} \\
\hline
\textbf{10-10-10-1} & $210$ & $ 2^{13.0}$ & $2^{-18.2}$\\
\hline
\textbf{10-20-20-1} & $620$ &$ 2^{14.5}$ & $2^{-17.8}$\\
\hline
\textbf{40-20-10-10-1} & $1,110$ &$ 2^{16.4}$ & $2^{-12.1}$\\
\hline
\textbf{80-40-20-1} & $4,020$  &$ 2^{19.1}$ & $2^{-14.8}$\\
\hline
\textbf{784-32-1} & $25,120$ &$ 2^{19.8}$ & $2^{-17.7}$ \\
\hline
\textbf{784-128-1} & $100,480$  &$ 2^{21.7}$ & $2^{-17.4}$ \\
\hline
\end{tabular}}
\end{table}

While our methodology achieves strong results, \autoref{tab:res_sota_64} and \autoref{tab:res_sota_supp_32} indicate that both the depth and the number of neurons per layer affect the attack's performance.
In the next section, we further analyze this limitation and evaluate the impact of special-case neurons, as well as the extraction of the last layer.

\section{Discussion of experimental part} \label{sec:extraction_32bits_discussion}

\subsection{Propagation of errors} \label{subsec:extraction_32bits_error}

\corr{In \autoref{subsec:divide_conquer_methodo}, we introduced our strategy as a divide-and-conquer approach, which requires knowledge of the $(i-1)$-th layer to extract the $i$-th one. 
Thus, one may question the impact of error propagation if the $(i-1)$-th layer is not successfully extracted.}
As mentioned in \autoref{subsec:extraction_32bits_split}, using binary search, it is only possible to place a critical point within a certain distance of the hyperplane.
This introduces a small error in the estimation of the weights, as illustrated in \autoref{tab:stat_layer}, which provides the layer-wise statistics of the shortened MobileNetv1 extraction. 
The propagation of this small error in the current layer's weights accumulates in the activation values, \textit{i.e.}, the inputs for the system of equations of the next layer, which reduces the extraction performance of subsequent layers.
This pattern is clearly visible in \autoref{tab:stat_layer}.
For the first layer L0, the maximum error on the estimated weights is $2^{-18.9}$, while for the last hidden layer, this error increases to $2^{-4.3}$.
For comparison, we also show results using $64$-bit data.
We observe the same pattern; the difference is that higher precision in localizing critical points reduces the error after the first layer to $2^{-46.6}$, compared to $32$-bit data, resulting in a lower final error of $2^{-15.3}$. 
This error propagation is not intrinsic to our method, as all SOTA cryptanalytic extraction methodologies are affected by it.
It is therefore crucial to minimize the error at each layer to reduce its impact on subsequent layers, highlighting the benefits of using an oracle-based methodology. 
Alternatively, several methods could be implemented to reduce this propagation, such as re-computing the weights to achieve higher precision. However, we leave this improvement for future work.

\begin{table*}
\caption{Layer-wise statistics of the extraction of the shortened MobileNetv1.}
\label{tab:stat_layer}
\centering
\resizebox{\textwidth}{!}{
\begin{tabular}{|p{0.12\linewidth} | p{0.13\linewidth}| p{0.04\linewidth}  p{0.04\linewidth}  p{0.04\linewidth} p{0.04\linewidth} p{0.04\linewidth} p{0.04\linewidth} p{0.04\linewidth}  p{0.04\linewidth}  p{0.04\linewidth} p{0.04\linewidth} p{0.04\linewidth} p{0.04\linewidth}|}
\hline
\textbf{Datatype} & \textbf{Metrics} & \textbf{L0} & \textbf{L1} & \textbf{L2} & \textbf{L3} & \textbf{L4} & \textbf{L5} & \textbf{L6} & \textbf{L7} & \textbf{L8} & \textbf{L9} & \textbf{L10} & \textbf{L11}  \\
\hline
$32$ bits & $\texttt{max}|\theta - \hat{\theta}|^{L}$ & $2^{-18.9}$ & $2^{-17.6}$ & $2^{-7.9}$ & $2^{-18.2}$ & $2^{-7.6}$ & $2^{-13.9}$ & $2^{-9.9}$ & $2^{-11.4}$ & $2^{-8.8}$ & $2^{-6.7}$ & $2^{-4.3}$ & $2^{3.7}$\\
 \hline
$64$ bits & $\texttt{max}|\theta - \hat{\theta}|^{L}$ & $2^{-46.6}$ & $2^{-43.8}$ & $2^{-37.4}$ & $2^{-34.2}$ & $2^{-29.0}$ & $2^{-27.1}$ & $2^{-26.0}$ & $2^{-26.8}$ & $2^{-23.1}$ & $2^{-22.7}$ & $2^{-15.3}$ & $2^{3.8}$\\
 \hline
\end{tabular}}
\end{table*}

\subsection{Impact of the special-case neurons} \label{subsec:extraction_32bits_special_cases}

\begin{figure}[ht]
    \centering
    \includegraphics[width=0.75\textwidth]{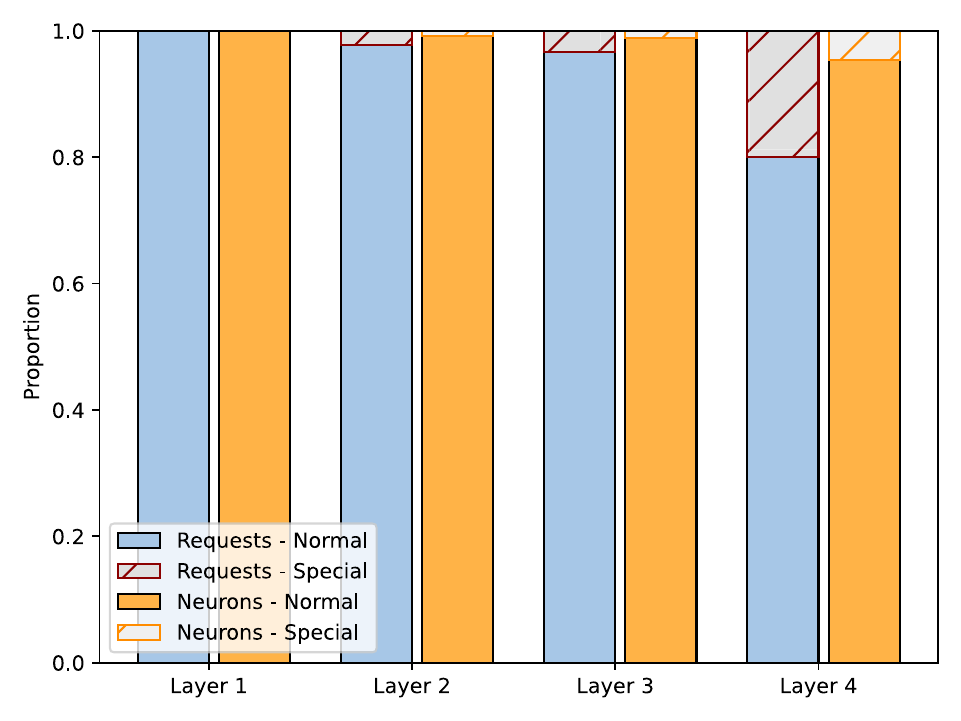}
    \caption{Comparison of the proportion of requests dedicated to normal versus special-case neurons.}
    \label{fig:impact_special_case}
\end{figure}

Another point of interest is the impact of special-case neurons.
Through the oracle, it is possible to improve the robustness of the extraction process, but their extraction remains costly in terms of queries.
To visualize their impact, we present the share of special-case neurons in the total number of requests and neurons for the 3072-256-256-256-64-10 MLP in \autoref{fig:impact_special_case}.
While previous works mention the existence of \textit{always-off} and \textit{always-on} neurons, their presence is not considered during extraction, and existing methodologies do not propose solutions for these neurons.
This is especially true for \textit{input-off} neurons, which are not even acknowledged in SOTA.
However, we observe such neurons in our experiments, even in simple architectures.
In particular, we compute the proportion of queries dedicated to special-case neurons per layer, as well as the proportion of these neurons per layer, as shown in \autoref{fig:impact_special_case} for the 3072-512-256-64-10 MLP.
While most of these special neurons correspond to \textit{input-off} and could theoretically be extracted by previous methodologies, doing so greatly increases the cost of the attack compared to methods using an oracle to optimize extraction. \\

These results highlight the impact of special-case neurons on the number of requests required to extract a DNN, even for a simple architecture such as a five-layer MLP.

\subsection{Extraction of the last layer} \label{subsec:extraction_32bits_last_layer}

As previously mentioned, extraction differs for the last layer. 
This is why we include both the maximum error on the first $L-1$ layers and on all $L$ layers, \textit{i.e.}, the full network.
Since there is no non-linear activation function for this layer, no critical points can be found.
In \cite{Chen2024HardLabelCE}, the authors leverage the fact that they target binary classifiers to precisely extract the last layer.
To improve the generality of our methodology, we instead use supervised learning on a dataset composed of the random inputs generated for the previous layers and their corresponding hard-labels to train the last layer to behave similarly to the targeted model's last layer.
While this methodology allows extraction of the last layer in hard-label settings, it accounts for most of the error in our process.

To validate this, we combine the first eleven extracted layers with the true final layer of the targeted model.
This hybrid model achieves a near-perfect fidelity of $99.56\%$, confirming that the main loss in fidelity comes from the last layer extraction.
As this limitation affects all methods based on extracting critical points in hard-label settings on multi-class classifiers, improving last-layer extraction is left for future work.
Two alternative methods could mitigate this issue.
The first is the methodology in \cite{Coqueret2023WhenSA}, performing a side-channel attack to retrieve confidence scores before extracting the last layer, which requires a profiling device and thus changes the threat model.
The second is the methodology in \cite{Canales_extraction_last_layer}, \textit{i.e.}, fixing several parameters and extracting the last layer under these constraints.
This approach could be used as a supplementary attack to improve final-layer fidelity.
However, while it achieves high label agreement, the last layer is not completely extracted, and its impact on downstream attacks, such as evasion attacks, has not been investigated \cite{Canales_extraction_last_layer}.

\subsection{\corr{Impact the error rate of the oracle}}\label{subsec:extraction_32bits_impact_error_rate}
\corr{In \autoref{subsec:extraction_32bits_result_classifiers} and \autoref{subsec:extraction_32bits_result_simulation}, we assume access to a perfect oracle, enabling the attacker to perfectly compare the states of two neurons with a single EM trace. 
However, as shown in \autoref{subsec:extraction_32bits_vulnerability}, this is not the case in practice, and the oracle may fail to compare neuron states. 
A single wrong estimation during binary search can produce an incorrectly estimated critical point, resulting in a biased system of equations that may deteriorate weight estimation or render the system unsolvable.
On average, placing one critical point in our setup requires 40 oracle calls.
If the oracle has an 85.9\% success rate, only 0.2\% of the estimated critical points are truly critical.
With a 99.9\% success rate, achievable in 13 EM traces, this percentage rises to 96.1\%.
This highlights the need for a near-perfect oracle with minimal side-channel traces.
We identify two complementary strategies: reduce the oracle's error rate per trace and make the extraction process robust to incorrect critical points.
The first depends heavily on the targeted device; attackers could improve the oracle by physically accessing the chip to improve signal quality or enhance side-channel trace preprocessing.
Although we use the k-means algorithm here, different clustering methods could further improve oracle accuracy.
For the second strategy, our methodology could be adapted to consider equations from critical points as correct with probability $p$, reflecting oracle uncertainty.
Under this adaptation, the attacker would sample extra critical points and solve the LSTSQ problem on subsamples, similarly to a RANSAC algorithm \cite{RANSAC}.
Validation of these adaptations is left for future work.}

\subsection{\corr{Countermeasures}}
\corr{Our methodology relies on the oracle $\mathsf{O}$, \textit{i.e.}, the side-channel attack on the ReLU function. 
If the attacker can no longer compare neuron states for specific inputs, our methodology becomes inapplicable.
Therefore, countermeasures should specifically protect activation functions.
While the constant-time implementation in \cite{Maji2021LeakyNR} exists, it remains vulnerable to other side-channel attacks, as demonstrated here.
One option is to operate on masked values or use a LUT-based ReLU implementation.
Another, without modifying the ReLU implementation, is to shuffle the neuron execution order at each run.
Our binary search, see \autoref{subsec:extraction_32bits_split}, only succeeds if the neuron execution order is known.
Randomizing the order makes binary search and critical point collection more difficult to be performed.
Evaluation of these countermeasures is left for future work.}
\section{Conclusion}
\label{conclusion}
In this study, we describe how side-channel attacks can be used as an oracle to improve fidelity-based model extraction attacks.
We design a new methodology inspired by cryptanalysis, introducing a new paradigm: divide and conquer.
With this approach, we demonstrate the extraction of new complex DNNs, including pooling layers, in hard-label settings while improving the performance of the extraction process.
Finally, we validate our approach by simulating the oracle and performing the extraction of the most complex architecture targeted in soft-label settings, showing that, despite a stricter threat model, our approach remains competitive and even outperforms previous methodologies in terms of accuracy on the extracted weights. \\

Several potential directions for future work can be considered to improve our methodology.
The first concerns the oracle.
While we have used unsupervised clustering on the EM traces corresponding to the processing of the ReLU function on an MCU, our oracle methodology is neither restricted to this algorithm, \textit{i.e.}, \textit{k-means}, nor to this activation function, nor to this type of physical device.
Therefore, our methodology could be extended to DNNs using any kind of activation function, provided it is possible to construct an oracle with the properties described in \autoref{subsec:extraction_32bits_split}.
Another challenge is to develop an oracle adapted to parallelized architectures such as GPUs, TPUs, or FPGAs.
As previously mentioned, in these systems the processing is parallelized, which may make side-channel leakages more difficult to exploit.
One possibility is to change the type of side-channel entirely by using time-to-digital converters, as in \cite{Lomet2025SideChannelEO}, or to consider alternative physical attacks and fault injection methods. 
The second concerns the adaptation of the methodology to other types of DNN architectures.
While in our experiments we target a MobileNetv1 and several MLPs, a large variety of architectures remain unexplored and could be considered under the assumption that a suitable oracle has been identified. \\

Finally, the most important perspective, from our point of view, is to study the capacity of our methodology to target quantized DNNs.
Indeed, in \autoref{subsec:extraction_32bits_error}, we describe how errors propagate from one layer to another and how the datatype used impacts the original error.
To the best of our knowledge, no previous work has studied the impact of quantization on cryptanalytic extraction, making this a promising direction for future work.
\appendix
\section{Description of the newly targeted architectures}\label{appendix:target_architecture}
An overall description of the targeted architecture based on MobileNetv1 is given in \autoref{tab:extraction32bits_archi_target_mobilenetv1} and of the depthwise separable convolution in \autoref{tab:archi_DW_sep}. 
Then in \autoref{tab:archi_bigMLP}, we detail the architecture of our large MLP.

\begin{table}[ht]
\caption{Description of the architecture of the shortened MobileNetv1.}
\label{tab:extraction32bits_archi_target_mobilenetv1}
\centering
\resizebox{\linewidth}{!}{
\begin{tabular}{|p{0.5\linewidth}|p{0.3\linewidth}| p{0.2\linewidth} |}
\hline
\textbf{Layers} & \textbf{Output size} & \textbf{Parameters}\\
\hline
Conv2D $3\times3$, padding, stride=$2$ & $1\times12\times16\times16$ & $324$\\
\hline
BatchNorm2D & $1\times12\times16\times16$ & $24$\\
\hline
ReLU & $1\times12\times16\times16$ & $0$\\
\hline
Depth-wise separable convolution & $1\times8\times16\times16$ & $244$\\
\hline
Depth-wise separable convolution & $1\times16\times8\times8$ & $248$\\
\hline
Depth-wise separable convolution & $1\times16\times8\times8$ & $464$\\
\hline
Depth-wise separable convolution & $1\times32\times4\times4$ & $752$\\
\hline
Depth-wise separable convolution & $1\times64\times4\times4$ & $2528$\\
\hline
Adaptive Average Pooling & $1\times64\times1\times1$ & $0$ \\
\hline
Dense & $1\times10$ & $650$ \\
\hline
\end{tabular}}
\end{table}

\begin{table}[ht]
\caption{Description of a depth-wise separable convolution taking $N_c$ channel as input.}
\label{tab:archi_DW_sep}
\centering
\resizebox{\linewidth}{!}{
\begin{tabular}{|p{0.3\linewidth}|p{0.2\linewidth}| p{0.2\linewidth} | p{0.2\linewidth} |p{0.1\linewidth} |}
\hline
\textbf{Layers} & \textbf{Kernel size} & \textbf{Padding} & \textbf{Grouped convolution} & \textbf{Bias}\\
\hline
Conv2D & $3\times3$ & 1 & \cmark & \xmark\\
BatchNorm2D &  &  &  & \\
ReLu & & & &\\
\hline
Conv2D & $N_c\times1\times1$ & 1 & \xmark & \xmark\\
BatchNorm2D &  &  &  & \\
ReLu & & & &\\
\hline
\end{tabular}}
\end{table}

\begin{table}[ht]
\caption{Description of the architecture of the large MLP.}
\label{tab:archi_bigMLP}
\centering
\resizebox{\linewidth}{!}{
\begin{tabular}{|p{0.5\linewidth}|p{0.3\linewidth}| p{0.2\linewidth} |}
\hline
\textbf{Layers} & \textbf{Output size} & \textbf{Parameters}\\
\hline
Flatten & $1\times3072$ & $0$\\
\hline
Dense + ReLU & $1\times512$ & $1,573,376$ \\
\hline
Dense + ReLU & $1\times256$ & $131,328$ \\
\hline
Dense + ReLU & $1\times64$ & $16,448$ \\
\hline
Dense + ReLU & $1\times10$ & $650$ \\
\hline
\end{tabular}}
\end{table}

\newpage
\bibliographystyle{alpha}
\bibliography{biblio}

@inproceedings{Jagielski2019HighAA,
  title={High Accuracy and High Fidelity Extraction of Neural Networks},
  author={Matthew Jagielski and Nicholas Carlini and David Berthelot and Alexey Kurakin and Nicolas Papernot},
  booktitle={USENIX Security Symposium},
  year={2019},
  url={https://api.semanticscholar.org/CorpusID:211858541}
}

@inproceedings{Rolnick2019ReverseengineeringDR,
  title={Reverse-engineering deep ReLU networks},
  author={David Rolnick and Konrad Paul Kording},
  booktitle={International Conference on Machine Learning},
  year={2019},
  url={https://api.semanticscholar.org/CorpusID:211259052}
}

@InProceedings{Shamir2023PolynomialTC,
author="Canales-Mart{\'i}nez, Isaac A.
and Ch{\'a}vez-Saab, Jorge
and Hambitzer, Anna
and Rodr{\'i}guez-Henr{\'i}quez, Francisco
and Satpute, Nitin
and Shamir, Adi",
editor="Joye, Marc
and Leander, Gregor",
title="Polynomial Time Cryptanalytic Extraction of Neural Network Models",
booktitle="Advances in Cryptology -- EUROCRYPT 2024",
year="2024",
publisher="Springer Nature Switzerland",
address="Cham",
pages="3--33",
isbn="978-3-031-58734-4"
}

@inproceedings{Carlini2020CryptanalyticEO,
  title={Cryptanalytic Extraction of Neural Network Models},
  author={Nicholas Carlini and Matthew Jagielski and Ilya Mironov},
  booktitle={Annual International Cryptology Conference},
  year={2020},
  url={https://api.semanticscholar.org/CorpusID:212644655}
}

@inproceedings{Joud2022API,
  title={A Practical Introduction to Side-Channel Extraction of Deep Neural Network Parameters},
  author={Raphael Joud and Pierre-Alain Mo{\"e}llic and Simon Ponti{\'e} and Jean-Baptiste Rigaud},
  booktitle={Smart Card Research and Advanced Application Conference},
  year={2022},
  url={https://api.semanticscholar.org/CorpusID:253447043}
}

@inproceedings{Hector2023FaultIA,
  TITLE = {{Fault injection and safe-error attack for extraction of embedded neural network models}},
  AUTHOR = {Hector, K{\'e}vin and Dumont, Mathieu and Moellic, Pierre-Alain and Dutertre, Jean-Max},
  URL = {https://cea.hal.science/cea-04607995},
  BOOKTITLE = {{ESORICS 2023 International Workshops}},
  ADDRESS = {La Hague, Netherlands},
  ORGANIZATION = {{Delft University of Technology}},
  SERIES = {Computer Security. ESORICS 2023 International Workshops CPS4CIP, ADIoT, SecAssure, WASP, TAURIN, PriST-AI, and SECAI, The Hague, The Netherlands, September 25--29, 2023, Revised Selected Papers, Part II},
  VOLUME = {14399},
  PAGES = {644-664},
  YEAR = {2023},
  MONTH = Sep,
  DOI = {10.1007/978-3-031-54129-2\_38},
  HAL_ID = {cea-04607995},
  HAL_VERSION = {v1},
}

@article{Rakin2021DeepStealAM,
  title={DeepSteal: Advanced Model Extractions Leveraging Efficient Weight Stealing in Memories},
  author={Adnan Siraj Rakin and Md Hafizul Islam Chowdhuryy and Fan Yao and Deliang Fan},
  journal={2022 IEEE Symposium on Security and Privacy (SP)},
  year={2021},
  pages={1157-1174},
  url={https://api.semanticscholar.org/CorpusID:243848074}
}

@inproceedings{horvath2025barracuda,
  title={$\{$BarraCUDA$\}$: Edge $\{$GPUs$\}$ do Leak $\{$DNN$\}$ Weights},
  author={Horvath, Peter and Chmielewski, Lukasz and Weissbart, L{\'e}o and Batina, Lejla and Yarom, Yuval},
  booktitle={34th USENIX Security Symposium (USENIX Security 25)},
  pages={4017--4034},
  year={2025}
}

@inproceedings{Batina2019CSINR,
  title={CSI NN: Reverse Engineering of Neural Network Architectures Through Electromagnetic Side Channel},
  author={Lejla Batina and Shivam Bhasin and Dirmanto Jap and Stjepan Picek},
  booktitle={USENIX Security Symposium},
  year={2019},
  url={https://api.semanticscholar.org/CorpusID:199558279}
}

@article{Zhang2023DeepLearningME,
  title={Deep-Learning Model Extraction Through Software-Based Power Side-Channel},
  author={Xiang Zhang and Aidong Adam Ding and Yunsi Fei},
  journal={2023 IEEE/ACM International Conference on Computer Aided Design (ICCAD)},
  year={2023},
  pages={1-9},
  url={https://api.semanticscholar.org/CorpusID:265525150}
}

@article{Milli2018ModelRF,
  title={Model Reconstruction from Model Explanations},
  author={Smitha Milli and Ludwig Schmidt and Anca D. Dragan and Moritz Hardt},
  journal={Proceedings of the Conference on Fairness, Accountability, and Transparency},
  year={2018},
  url={https://api.semanticscholar.org/CorpusID:49741763}
}

@inproceedings{Hanin2019DeepRN,
  title={Deep ReLU Networks Have Surprisingly Few Activation Patterns},
  author={Boris Hanin and David Rolnick},
  booktitle={Neural Information Processing Systems},
  year={2019},
  url={https://api.semanticscholar.org/CorpusID:173991223}
}

@article{Fredrikson2015ModelIA,
  title={Model Inversion Attacks that Exploit Confidence Information and Basic Countermeasures},
  author={Matt Fredrikson and Somesh Jha and Thomas Ristenpart},
  journal={Proceedings of the 22nd ACM SIGSAC Conference on Computer and Communications Security},
  year={2015}
}

@inproceedings{Szegedy2013IntriguingPO,
    title	= {Intriguing properties of neural networks},
    author	= {Christian Szegedy and Wojciech Zaremba and Ilya Sutskever and Joan Bruna and Dumitru Erhan and Ian Goodfellow and Rob Fergus},
    year	= {2014},
    URL	= {http://arxiv.org/abs/1312.6199},
    booktitle	= {International Conference on Learning Representations}
}

@article{Carlini2016TowardsET,
  title={Towards Evaluating the Robustness of Neural Networks},
  author={Nicholas Carlini and David A. Wagner},
  journal={2017 IEEE Symposium on Security and Privacy (SP)},
  year={2016},
  pages={39-57}
}

@ARTICLE{Dying_ReLU,
       author = {{Lu}, Lu},
        title = "{Dying ReLU and Initialization: Theory and Numerical Examples}",
      journal = {Communications in Computational Physics},
         year = 2020,
        month = jun,
       volume = {28},
       number = {5},
        pages = {1671-1706},
          doi = {10.4208/cicp.OA-2020-0165},
archivePrefix = {arXiv},
       eprint = {1903.06733},
 primaryClass = {stat.ML},
       adsurl = {https://ui.adsabs.harvard.edu/abs/2020CCoPh..28.1671L}
}

@INPROCEEDINGS{Gao2023DeepTheftSD,
  author={Gao, Yansong and Qiu, Huming and Zhang, Zhi and Wang, Binghui and Ma, Hua and Abuadbba, Alsharif and Xue, Minhui and Fu, Anmin and Nepal, Surya},
  booktitle={2024 IEEE Symposium on Security and Privacy (SP)}, 
  title={DeepTheft: Stealing DNN Model Architectures through Power Side Channel}, 
  year={2024},
  volume={},
  number={},
  pages={3311-3326},
  doi={10.1109/SP54263.2024.00250}}

@INPROCEEDINGS{9300274,
  author={Yu, Honggang and Ma, Haocheng and Yang, Kaichen and Zhao, Yiqiang and Jin, Yier},
  booktitle={2020 IEEE International Symposium on Hardware Oriented Security and Trust (HOST)}, 
  title={DeepEM: Deep Neural Networks Model Recovery through EM Side-Channel Information Leakage}, 
  year={2020},
  volume={},
  number={},
  pages={209-218},
  doi={10.1109/HOST45689.2020.9300274}}

@inproceedings{Cagli2017ConvolutionalNN,
  title={Convolutional Neural Networks with Data Augmentation Against Jitter-Based Countermeasures - Profiling Attacks Without Pre-processing},
  author={Eleonora Cagli and C{\'e}cile Canovas and Emmanuel Prouff},
  booktitle={Workshop on Cryptographic Hardware and Embedded Systems},
  year={2017},
  url={https://api.semanticscholar.org/CorpusID:54088207}
}

@inproceedings{inproceedings,
  title={Breaking cryptographic implementations using deep learning techniques},
  author={Maghrebi, Houssem and Portigliatti, Thibault and Prouff, Emmanuel},
  booktitle={Security, Privacy, and Applied Cryptography Engineering: 6th International Conference, SPACE 2016, Hyderabad, India, December 14-18, 2016, Proceedings 6},
  pages={3--26},
  year={2016},
  organization={Springer}
}

@inproceedings{Maia2021CanOH,
  title={Can one hear the shape of a neural network?: Snooping the GPU via Magnetic Side Channel},
  author={Henrique Teles Maia and Chang Xiao and Dingzeyu Li and Eitan Grinspun and Changxi Zheng},
  booktitle={USENIX Security Symposium},
  year={2021},
  url={https://api.semanticscholar.org/CorpusID:237513609}
}

@inbook{kmeanbook,
author = {Duda, Richard and Hart, Peter and G.Stork, David},
year = {2001},
month = {01},
pages = {},
title = {Pattern Classification},
isbn = {0-471-05669-3},
journal = {Wiley Interscience}
}

@article{Maji2021LeakyNR,
  title={Leaky Nets: Recovering Embedded Neural Network Models and Inputs Through Simple Power and Timing Side-Channels—Attacks and Defenses},
  author={Saurav Maji and Utsav Banerjee and Anantha P. Chandrakasan},
  journal={IEEE Internet of Things Journal},
  year={2021},
  volume={8},
  pages={12079-12092},
  url={https://api.semanticscholar.org/CorpusID:232404206}
}

@misc{Canales_extraction_last_layer,
  author = {Isaac A. Canales-Martínez and David Santos},
  title = {Extracting Some Layers of Deep Neural Networks in the Hard-Label Setting},
  howpublished = {Cryptology {ePrint} Archive, Paper 2025/1118},
  year = {2025},
  url = {https://eprint.iacr.org/2025/1118}
}

@article{Howard2017MobileNetsEC,
  title={MobileNets: Efficient Convolutional Neural Networks for Mobile Vision Applications},
  author={Andrew G. Howard and Menglong Zhu and Bo Chen and Dmitry Kalenichenko and Weijun Wang and Tobias Weyand and Marco Andreetto and Hartwig Adam},
  journal={ArXiv},
  year={2017},
  volume={abs/1704.04861},
  url={https://api.semanticscholar.org/CorpusID:12670695}
}

@InProceedings{Carlini2024PolynomialTC,
author="Carlini, Nicholas
and Ch{\'a}vez-Saab, Jorge
and Hambitzer, Anna
and Rodr{\'i}guez-Henr{\'i}quez, Francisco
and Shamir, Adi",
editor="Fehr, Serge
and Fouque, Pierre-Alain",
title="Polynomial Time Cryptanalytic Extraction of Deep Neural Networks in the Hard-Label Setting",
booktitle="Advances in Cryptology -- EUROCRYPT 2025",
year="2025",
publisher="Springer Nature Switzerland",
address="Cham",
pages="364--396",
isbn="978-3-031-91107-1"
}

@inproceedings{Joud2023LikeAO,
  author       = {Rapha{\"{e}}l Joud and
                  Pierre{-}Alain Mo{\"{e}}llic and
                  Simon Ponti{\'{e}} and
                  Jean{-}Baptiste Rigaud},
  editor       = {Shivam Bhasin and
                  Thomas Roche},
  title        = {Like an Open Book? Read Neural Network Architecture with Simple Power
                  Analysis on 32-Bit Microcontrollers},
  booktitle    = {Smart Card Research and Advanced Applications - 22nd International
                  Conference, {CARDIS} 2023, Amsterdam, The Netherlands, November 14-16,
                  2023, Revised Selected Papers},
  series       = {Lecture Notes in Computer Science},
  volume       = {14530},
  pages        = {256--276},
  publisher    = {Springer},
  year         = {2023},
  url          = {https://doi.org/10.1007/978-3-031-54409-5\_13},
  doi          = {10.1007/978-3-031-54409-5\_13},
  bibsource    = {dblp computer science bibliography, https://dblp.org}
}

@InProceedings{Chen2024HardLabelCE,
author="Chen, Yi
and Dong, Xiaoyang
and Guo, Jian
and Shen, Yantian
and Wang, Anyu
and Wang, Xiaoyun",
editor="Chung, Kai-Min
and Sasaki, Yu",
title="Hard-Label Cryptanalytic Extraction of Neural Network Models",
booktitle="Advances in Cryptology -- ASIACRYPT 2024",
year="2025",
publisher="Springer Nature Singapore",
address="Singapore",
pages="207--236",
isbn="978-981-96-0944-4"
}

@article{Liu2025NavigatingTD,
  title={Navigating the Deep: Signature Extraction on Deep Neural Networks},
  author={Haolin Liu and Adrien Siproudhis and Samuel Experton and Peter Lorenz and Christina Boura and Thomas Peyrin},
  journal={ArXiv},
  year={2025},
  volume={abs/2506.17047},
  url={https://api.semanticscholar.org/CorpusID:279465043}
}

@inproceedings{Coqueret2023WhenSA,
  title={When Side-Channel Attacks Break the Black-Box Property of Embedded Artificial Intelligence},
  author={Beno{\^i}t Coqueret and Mathieu Carbone and Olivier Sentieys and Gabriel Zaid},
  booktitle={Proceedings of the 16th ACM Workshop on Artificial Intelligence and Security},
  year={2023},
  url={https://api.semanticscholar.org/CorpusID:265360074}
}

@INPROCEEDINGS{Lomet2025SideChannelEO,
  author={Lomet, Guillaume and Salvador, Rubén and Colombier, Brice and Grosso, Vincent and Sentieys, Olivier and Killian, Cédric},
  booktitle={2025 IEEE 31st International Symposium on On-Line Testing and Robust System Design (IOLTS)}, 
  title={Side-Channel Extraction of Dataflow AI Accelerator Hardware Parameters}, 
  year={2025},
  volume={},
  number={},
  pages={1-7},
  doi={10.1109/IOLTS65288.2025.11117043}
}

@misc{Ito2025HardLabelExtraction,
      author = {Akira Ito and Takayuki Miura and Yosuke Todo},
      title = {Is the Hard-Label Cryptanalytic Model Extraction Really Polynomial?},
      howpublished = {Cryptology {ePrint} Archive, Paper 2025/1868},
      year = {2025},
      url = {https://eprint.iacr.org/2025/1868}
}

@inproceedings{
madry2018towards,
title={Towards Deep Learning Models Resistant to Adversarial Attacks},
author={Aleksander Madry and Aleksandar Makelov and Ludwig Schmidt and Dimitris Tsipras and Adrian Vladu},
booktitle={International Conference on Learning Representations},
year={2018},
url={https://openreview.net/forum?id=rJzIBfZAb},
}

@inproceedings{carlini2017towards,
  title={Towards evaluating the robustness of neural networks},
  author={Carlini, Nicholas and Wagner, David},
  booktitle={2017 ieee symposium on security and privacy (sp)},
  pages={39--57},
  year={2017},
  organization={Ieee}
}

@incollection{kurakin2018adversarial,
  title={Adversarial examples in the physical world},
  author={Kurakin, Alexey and Goodfellow, Ian J and Bengio, Samy},
  booktitle={Artificial intelligence safety and security},
  pages={99--112},
  year={2018},
  publisher={Chapman and Hall/CRC}
}

@article{papernot2018cleverhans,
  title={Technical Report on the CleverHans v2.1.0 Adversarial Examples Library},
  author={Nicolas Papernot and Fartash Faghri and Nicholas Carlini and
  Ian Goodfellow and Reuben Feinman and Alexey Kurakin and Cihang Xie and
  Yash Sharma and Tom Brown and Aurko Roy and Alexander Matyasko and
  Vahid Behzadan and Karen Hambardzumyan and Zhishuai Zhang and
  Yi-Lin Juang and Zhi Li and Ryan Sheatsley and Abhibhav Garg and
  Jonathan Uesato and Willi Gierke and Yinpeng Dong and David Berthelot and
  Paul Hendricks and Jonas Rauber and Rujun Long},
  journal={arXiv preprint arXiv:1610.00768},
  year={2018}
}

@article{RANSAC,
author = {Fischler, Martin A. and Bolles, Robert C.},
title = {Random sample consensus: a paradigm for model fitting with applications to image analysis and automated cartography},
year = {1981},
issue_date = {June 1981},
publisher = {Association for Computing Machinery},
address = {New York, NY, USA},
volume = {24},
number = {6},
issn = {0001-0782},
url = {https://doi.org/10.1145/358669.358692},
doi = {10.1145/358669.358692},
journal = {Commun. ACM},
month = jun,
pages = {381–395},
numpages = {15}
}

\end{document}